\documentclass[]{aastex63}
\usepackage{ulem}

\newcommand\subs[1]{\textsubscript{#1}}
\newcommand\sups[1]{\textsuperscript{#1}}
\newcommand\rh[1]{\textcolor{black}{{\textit{r}\subs{H}}#1}}
\newcommand\lp[1]{\textcolor{black}{{\textit{L}\subs{p}}#1}}

\newcommand\qco[1]{\textcolor{black}{{\textit{Q}\subs{CO}}#1}}

\newcommand\trot[1]{\textcolor{black}{{\textit{T}\subs{rot}}#1}}
\newcommand\Ju[1]{\textcolor{black}{{\textit{J}}#1}}
\newcommand\ie[1]{\textcolor{black}{{\textit{i.e.,}}#1}}
\newcommand\eg[1]{\textcolor{black}{{\textit{e.g.,}}#1}}
\newcommand\kms[1]{\textcolor{black}{{km\,s$^{-1}$}#1}}

\received{April 21, 2023}
\revised{}
\accepted{}
\submitjournal{PSJ}

\shorttitle{}
\shortauthors{Roth et al.}

\begin{document}

\title{Molecular Outgassing in Centaur 29P/Schwassmann-Wachmann 1 During Its Exceptional 2021 Outburst: Coordinated Multi-Wavelength Observations Using nFLASH at APEX and iSHELL at the NASA-IRTF}

\correspondingauthor{Nathan X. Roth}
\email{nathaniel.x.roth@nasa.gov}

\author[0000-0002-6006-9574]{Nathan X. Roth}
\altaffiliation{Visiting Astronomer at the Infrared Telescope Facility, which is operated by the University of Hawaii under contract NNH14CK55B with the National Aeronautics and Space Administration.}
\affiliation{Solar System Exploration Division, Astrochemistry Laboratory Code 691, NASA Goddard Space Flight Center, 8800 Greenbelt Rd., Greenbelt, MD 20771, USA}
\affiliation{Department of Physics, The Catholic University of America, 620 Michigan Ave., N.E. Washington, DC 20064, USA}

\author [0000-0001-7694-4129]{Stefanie N. Milam}
\altaffiliation{Visiting Astronomer at the Infrared Telescope Facility, which is operated by the University of Hawaii under contract NNH14CK55B with the National Aeronautics and Space Administration.}
\affiliation{Solar System Exploration Division, Astrochemistry Laboratory Code 691, NASA Goddard Space Flight Center, 8800 Greenbelt Rd., Greenbelt, MD 20771, USA}

\author[0000-0001-8843-7511]{Michael A. DiSanti}
\altaffiliation{Visiting Astronomer at the Infrared Telescope Facility, which is operated by the University of Hawaii under contract NNH14CK55B with the National Aeronautics and Space Administration.}
\affiliation{Solar System Exploration Division, Planetary Systems Laboratory Code 693, NASA Goddard Space Flight Center, 8800 Greenbelt Rd., Greenbelt, MD 20771, USA}

\author[0000-0002-2662-5776]{Geronimo L. Villanueva}
\altaffiliation{Visiting Astronomer at the Infrared Telescope Facility, which is operated by the University of Hawaii under contract NNH14CK55B with the National Aeronautics and Space Administration.}
\affiliation{Solar System Exploration Division, Planetary Systems Laboratory Code 693, NASA Goddard Space Flight Center, 8800 Greenbelt Rd., Greenbelt, MD 20771, USA}

\author[0000-0003-0194-5615]{Sara Faggi}
\altaffiliation{Visiting Astronomer at the Infrared Telescope Facility, which is operated by the University of Hawaii under contract NNH14CK55B with the National Aeronautics and Space Administration.}
\affiliation{Solar System Exploration Division, Planetary Systems Laboratory Code 693, NASA Goddard Space Flight Center, 8800 Greenbelt Rd., Greenbelt, MD 20771, USA}
\affiliation{Department of Physics, American University, 4400 Massachusetts Avenue NW, Washington, D.C., 20016 USA}

\author[0000-0002-6391-4817]{Boncho P. Bonev}
\affiliation{Department of Physics, American University, 4400 Massachusetts Avenue NW, Washington, D.C., 20016 USA}

\author[0000-0001-8233-2436]{Martin A. Cordiner}
\affiliation{Solar System Exploration Division, Astrochemistry Laboratory Code 691, NASA Goddard Space Flight Center, 8800 Greenbelt Rd., Greenbelt, MD 20771, USA}
\affiliation{Department of Physics, The Catholic University of America, 620 Michigan Ave., N.E. Washington, DC 20064, USA}

\author[0000-0001-9479-9287]{Anthony J. Remijan}
\affiliation{National Radio Astronomy Observatory, 520 Edgemont Rd, Charlottesville, VA 22903, USA}

\author{Dominique Bockelée-Morvan}
\affiliation{LESIA, Observatoire de Paris, Université PSL, CNRS, Sorbonne Université,
Université de Paris, 5 place Jules Janssen, F-92195 Meudon, France}

\author[0000-0003-2414-5370]{Nicolas Biver}
\affiliation{LESIA, Observatoire de Paris, Université PSL, CNRS, Sorbonne Université,
Université de Paris, 5 place Jules Janssen, F-92195 Meudon, France}

\author{Jacques Crovisier}
\affiliation{LESIA, Observatoire de Paris, Université PSL, CNRS, Sorbonne Université,
Université de Paris, 5 place Jules Janssen, F-92195 Meudon, France}

\author[0000-0002-0500-4700]{Dariusz C. Lis}
\affiliation{Jet Propulsion Laboratory, California Institute of Technology, 4800 Oak Grove Drive, Pasadena, CA 91109, USA}

\author[0000-0001-6752-5109]{Steven B. Charnley}
\affiliation{Solar System Exploration Division, Astrochemistry Laboratory Code 691, NASA Goddard Space Flight Center, 8800 Greenbelt Rd., Greenbelt, MD 20771, USA}

\author{Emmanuel Jehin}
\affiliation{Space Sciences, Technologies \& Astrophysics Research (STAR) Institute, University of Liège, Belgium}

\author[0000-0002-0656-876X]{Eva S. Wirstr\"{o}m}
\affiliation{Department of Space, Earth and Environment, Chalmers University of Technology, Onsala Space Observatory, SE-439 92 Onsala, Sweden}

\author[0000-0002-0622-2400]{Adam J. McKay}
\altaffiliation{Visiting Astronomer at the Infrared Telescope Facility, which is operated by the University of Hawaii under contract NNH14CK55B with the National Aeronautics and Space Administration.}
\affiliation{Department of Physics \& Astronomy, Appalachian State University, 525 Rivers Street, Boone, NC 28608}



\begin{abstract}

The extraordinary 2021 September--October outburst of Centaur 29P/Schwassmann–Wachmann 1 afforded an opportunity to test the composition of primitive Kuiper disk material at high sensitivity. We conducted nearly simultaneous multi-wavelength spectroscopic observations of 29P/Schwassmann–Wachmann 1 using iSHELL at the NASA Infrared Telescope Facility and nFLASH at the Atacama Pathfinder EXperiment (APEX) on 2021 October 6, with follow-up APEX/nFLASH observations on 2021 October 7 and 2022 April 3. This coordinated campaign between near-infrared and radio wavelengths enabled us to sample molecular emission from a wealth of coma molecules and to perform measurements that cannot be accomplished at either wavelength alone. We securely detected CO emission on all dates with both facilities, including velocity-resolved spectra of the CO (\Ju{}=2--1) transition with APEX/nFLASH and multiple CO ($v$=1--0) rovibrational transitions with IRTF/iSHELL. We report rotational temperatures, coma kinematics, and production rates for CO and stringent (3$\sigma$) upper limits on abundance ratios relative to CO for CH$_4$, C$_2$H$_6$, CH$_3$OH, H$_2$CO, CS, and OCS. Our upper limits for CS/CO and OCS/CO represent their first values in the literature for this Centaur. Upper limits for CH$_4$, C$_2$H$_6$, CH$_3$OH, and H$_2$CO are the most stringent reported to date, and are most similar to values found in ultra CO-rich Oort cloud comet C/2016 R2 (PanSTARRS), which may have implications for how ices are preserved in cometary nuclei. We demonstrate the superb synergy of coordinated radio and near-infrared measurements, and advocate for future small body studies that jointly leverage the capabilities of each wavelength.

\end{abstract}



\section{Introduction} \label{sec:intro}
The study of comets affords a unique window into the birth, infancy, and subsequent evolution of the solar system. Soon after their accretion from the protosolar disk at the time of planet formation, comets were gravitationally scattered across the solar system, with many emplaced in their present-day dynamical reservoirs, the Oort cloud or the Kuiper disk, where they have remained in the cold outer solar system for the last $\sim$4.5 Gyr, affected by
minimal thermal and radiative processing. Systematically characterizing the compositions of their nuclei should therefore provide insights into the composition and thermochemical processes in the solar nebula where (and when) they formed \citep{Bockelee2004,Mumma2011a,DelloRusso2016a,Bockelee2017}. 

Comets that become available for remote sensing can be broadly categorized into two groups based on their dynamical reservoir: (1) The Jupiter-family comets (JFCs), originating from the scattered Kuiper disk with inclinations largely within the ecliptic plane, and (2) nearly isotropic Oort cloud comets (OCCs) stored in the far outer reaches of the solar system with random orbital inclinations. While these represent the major dynamical reservoirs for comets, the line between them has become increasingly ambiguous as ever more small bodies are discovered with the increased sensitivity and coverage of all-sky surveys. 

Although most processes thought to affect nucleus compositon during a typical perihelion passage should only penetrate the uppermost few meters \citep[see][]{Stern2003}, a JFC which experiences many perihelion passages (perhaps at small heliocentric distances, \rh{}) may undergo considerable thermal processing compared to an OCC, whereas an OCC may experience greater cosmic ray processing \citep{Harrington2022}. Indeed, there is evidence that JFCs are depleted in certain volatiles, such as acetylene (C$_2$H$_2$) and ethane (C$_2$H$_6$), with respect to OCCs \citep{DelloRusso2016a}. On the other hand, some JFCs have been shown to have abundances of the hypervolatile methane (CH$_4$) consistent with that seen in OCCs \citep{DiSanti2017,Roth2020}, and optical wavelength studies show no correlation between carbon-chain depletion and dynamical age \citep{AHearn1995}, suggesting that observed compositional differences between the two classes are not evolutionary. Understanding to what extent present-day measurements of coma abundances reflect natal and/or evolved chemistry in the nucleus is critical to placing these observations into the context of solar system formation.

The enigmatic objects known as Centaurs provide an opportunity to disentangle signatures imprinted in cometary ices by the nascent solar system from those acquired through thermal or other evolutionary processing. Centaurs are thought to be transition objects, migrating from the scattered Kuiper disk to their ultimate dynamical fate as JFCs \citep{Jewitt2009}. With semimajor axes and perihelia confined between those of Jupiter and Neptune, these distantly active objects have undergone considerably less potential thermal processing than their evolved JFC counterparts. As such, active Centaurs (those showing evidence of gas or dust comae) represent some of the most primitive Kuiper disk material available for study via remote sensing. Characterizing their volatile composition may serve as a bridge between the more primitive OCCs and the more processed JFCs. With advances in state-of-the-art observatories, Centaurs have come under increased study at both ground- and space-based facilities, including the James Webb Space Telescope (JWST) and the Atacama Large Millimeter/submillimeter Array (ALMA), and the recent 2023-2032 Planetary Science \& Astrobiology Decadal Survey recommended a Centaur orbiter and lander as one response to the New Fronters 6 call \citep{2023Decadal}.

Centaur 29P/Schwassmann-Wachmann 1 (hereafter SW1) is among the most active known Centaurs, with an orbital period \textit{P} = 14.65 years and a roughly spherical nucleus of radius 31 km \citep{Bockelee2022}. Its most recent perihelion was on 2019 March 7 at \textit{q} = 5.72 au. Documentation of its repeated outbursts dates back nearly a century \citep{Hughes1990}, and its peculiar cycles of quiescent activity punctuated by outbursts are the subject of significant study at multiple wavelengths stretching into the modern era \citep[e.g.,][]{Jewitt2009,Paganini2013,Wierzchos2020,Bockelee2022}. In terms of volatiles, carbon monoxide (CO), hydrogen cyanide (HCN) and water (H$_2$O) have been securely detected \citep{Ootsubo2012,Bockelee2022}, along with product species CO$^+$, the cyano radical (CN), and ionized dinitrogen (N$_2^+$) \citep{Cochran1991,Korsun2008,Ivanova2016}. Long term monitoring has revealed that dust and gas production are not always correlated during outbursts \citep{Wierzchos2020}. The drivers of activity at such large \rh{} ($\sim$6 au) beyond the H$_2$O sublimation zone are still under debate, although the transition of amorphous-to-crystalline water ice is a strong candidate \citep[\eg{}][]{Prialnik1987,Jewitt2009,Meech2009}.

Here we report coordinated multi-wavelength observations of SW1 taken during its exceptional late-2021 outburst \citep{Miles2021} using iSHELL at the NASA Infrared Telescope Facility (IRTF) and nFLASH230 at the Atacama Pathfinder EXperiment (APEX). The outburst began on September 25, with SW1 brightening nearly 6 magnitudes by September 28. SW1 slowly approached quiescent magnitudes over the next month, before undergoing a smaller outburst on October 23. IRTF/iSHELL and APEX/nFLASH230 observations were taken contemporaneously on 2021 October 6, with follow-up APEX/nFLASH230 observations on 2021 October 7 and 2022 April 3. Our April observations were designed to serve as a baseline taken outside of the outburst for comparison with the October observations. 

Secure detections of CO emission were identified at both wavelengths, and stringent (3$\sigma$) upper limits were determined for CH$_4$, C$_2$H$_6$, methanol (CH$_3$OH), formaldehyde (H$_2$CO), carbonyl sulfide (OCS), and carbon monosulfide (CS). We present molecular production rates, spectral line profiles, spatial profiles of column density, and abundance ratios (relative to CO). Section~\ref{sec:obs} discusses the observations. Sections~\ref{sec:apex-results} and~\ref{sec:ishell-results} detail our data reduction and present our results. Section~\ref{sec:discussion} provides optical context for the outburst, compares our results against previous radio wavelength measurements of SW1, and places them into the context of the molecular abundances measured in the larger comet population. 

\section{Observations} \label{sec:obs}
We conducted multi-wavelength observations to characterize SW1's chemistry during its brightest outburst in a decade. We chose near-infrared and millimeter wavelength observations for their highly complementary nature: APEX/nFLASH230 sampled CO (\Ju{}=2--1), H$_2$CO ($J_{Ka,Kc}$=$3_{0,3}$--$2_{0,2}$, $J_{Ka,Kc}$=$3_{1,2}$--$2_{1,1}$), CH$_3$OH ($J_K$=$5_0$--$4_0$ $A^+$), and CS (\Ju{}=5--4) emissions at high spectral resolution to derive detailed coma kinematics, and IRTF sampled multiple ro-vibrational transitions of CO to measure coma rotational temperature while also characterizing abundances of the symmetric hydrocarbons CH$_4$ and C$_2$H$_6$, along with CH$_3$OH and OCS. Table~\ref{tab:obslog} provides an observing log, and we detail our observations and data reduction procedures for each observatory in turn.

\begin{deluxetable*}{ccccccccccc}
\tablenum{1}
\tablecaption{Observing Log\label{tab:obslog}}
\tablewidth{0pt}
\tablehead{
\colhead{Date} & \colhead{UT Time} & \colhead{Setting} &\colhead{Target} & \colhead{\textit{T}\subs{int}} & \colhead{\textit{r}\subs{H}} & \colhead{$\Delta$} & \colhead{d$\Delta$/dt} & \colhead{Molecules} & \colhead{Slit PA} \\
\colhead{} & \colhead{} & \colhead{} & \colhead{} & \colhead{(min)}  & \colhead{(au)} & \colhead{(au)} & \colhead{(\kms{})} & \colhead{Sampled} & \colhead{($\degr$)}
}
\startdata
\multicolumn{10}{c}{IRTF/iSHELL} \\
\hline
2021 Oct 6 & 10:46 -- 12:41 & M2 & SW1 & 78 & 5.91 & 5.41 & -24.0 & CO, H$_2$O, OCS & 259 \\
& 12:56 -- 13:02 & M2 & BS-1641 & -- & -- & -- & -- & -- & -- \\
& 13:18 -- 13:21 & Lp1 & BS-1641 & -- & -- & -- & -- & -- & -- \\
& 13:28 -- 15:37 & Lp1 & SW1 & 108 & 5.91 & 5.41 & -23.7 & CH$_4$, C$_2$H$_6$, CH$_3$OH, H$_2$CO & 259 \\
\hline
\multicolumn{10}{c}{APEX/nFLASH230} \\
\hline
Date & UT Time & Setting & Target & \textit{T}\subs{int} & \rh{} & $\Delta$ & $\nu$ & Molecules & $\theta$ \\
& & & & (min) & (au) & (au) & (GHz) & Sampled & ($\arcsec$) \\
\hline
2021 Oct 6 & 06:12 - 10:23 & 1 & SW1 & 75 & 5.91 & 5.41 & 230.5 & CO, H$_2$CO, CH$_3$OH, CS & 26.2 \\
2021 Oct 7 & 06:48 - 10:01 & 1 & SW1 & 54 & 5.91 & 5.39 & 230.5 & CO, H$_2$CO, CH$_3$OH, CS & 26.2 \\
2022 Apr 3 & 18:15 - 20:15 & 1 & SW1 & 33 & 5.97 & 6.43 & 230.5 & CO, H$_2$CO, CH$_3$OH, CS & 26.2 \\
\enddata
\tablecomments{\textbf{IRTF/iSHELL Observations.} \textit{r}\subs{H}, $\Delta$, and d$\Delta$/dt are the heliocentric distance, geocentric distance, and geocentric velocity, respectively, of SW1 at the time of observations. \textit{T}\subs{int} is the integrated time on-source. The seeing on Maunakea varied from $\sim$0$\farcs$6--1$\farcs$1 and the average precipitable water vapor (PWV) was 1.2 mm. \textbf{APEX/nFLASH230 Observations.} $\theta$ is the primary beam size at $\nu$, the center frequency of the band. The average PWV was 5 mm, 4.5 mm, and 1.5 mm, and the solar phase angle (Sun-comet-Earth) was 8.7$^{\circ}$, 8.6$^{\circ}$, and 8.2$^{\circ}$ on 2021 October 6 and 7 and 2022 April 3, respectively.}
\end{deluxetable*}

\subsection{IRTF/\lowercase{i}SHELL Observations} \label{subsec:iSHELL}
We conducted spectroscopic observations of SW1 using the high-resolution, near-infrared facility spectrograph iSHELL \citep{Rayner2012,Rayner2016} at the 3 m NASA-IRTF on 2021 October 6 (Table~\ref{tab:obslog}). We utilized two iSHELL settings so as to efficiently sample a suite of molecular abundances: Lp1 samples CH$_4$, C$_2$H$_6$, H$_2$CO, and CH$_3$OH transitions near 3.3 - 3.5 $\mu$m, and M2 samples CO, OCS, and H$_2$O near 4.5 $\mu$m. We oriented the slit along the projected Sun-comet line (259$^{\circ}$). 

Observations were performed with a 6-pixel (0$\farcs$75) wide slit with resolving power ($\lambda/\Delta\lambda) \sim 4.5\times10^4$. We used a standard ABBA nod pattern in which the telescope is nodded along the slit between successive exposures, thereby placing the comet at two distinct positions along the slit (``A'' and ``B'') in order to facilitate sky subtraction. The A and B beams were symmetrically placed about the midpoint along the 15$\arcsec$ long slit and separated by half its length. SW1 was bright and easily acquired with IRTF/iSHELL's near-infrared active guiding system using the \Ju{}-band filter. Combining spectra of the nodded beams as A-B-B$+$A canceled emissions from thermal background, instrumental biases, and sky emission (lines and continuum) to second order in air mass. Flux calibration was performed using an appropriately placed bright infrared flux standard star (BS-1641) using a wide (4$\farcs$0) slit. 

\subsection{APEX/\lowercase{n}FLASH230 Observations} \label{subsec:APEX}
We conducted single dish, position-switched observations of SW1 using APEX \citep{Gusten2006} on 2021 October 6 and 7 and 2022 April 3 using the nFLASH230 receiver with the Fast Fourier Transform Spectrometer (FFTS) backends, covering frequencies between 210.77 and 247.57 GHz ($\lambda$ = 1.21 -- 1.42 mm) at a frequency resolution of 61 kHz ($\sim$0.08 \kms{}). Offset spectra were taken at a distance of 180$\arcsec$ from the position of SW1. The APEX/nFLASH230 beam full width at half maximum (FWHM) at these frequencies ranges from 24$\farcs$3 -- 28$\farcs$6, corresponding to nucleocentric distances of 95,000 -- 112,000 km and 113,000 -- 133,000 km, respectively, at the geocentric distance of SW1 in October (5.4 au) and in April (6.4 au). SW1's position was tracked using JPL HORIZONS ephemerides (JPL \#K192/71). The weather (average precipitable water vapor at zenith, PWV) was fair (5 mm, 4.5 mm, and 1.5 mm, on October 6, 7, and April 3, respectively). Pointing and focus scans were obtained with the Mira variable o Ceti. Flux calibration scans were carried out regularly throughout the night using o Ceti and the evolved star CRL 618. We reduced the data using the \texttt{GILDAS/CLASS} software package\footnote{\url{http://www.iram.fr/IRAMFR/GILDAS}}. We subtracted first-order polynomial fits to line-free spectral regions near each targeted transition to remove the continuum, converted the spectra to velocity space in the cometocentric rest frame, and placed the fluxes onto the main beam scale using main beam efficiencies cataloged on the APEX website\footnote{https://www.apex-telescope.org/telescope/efficiency/} near our observation dates: $\eta_{MB}$ = 0.78 and 0.81 in October and April, respectively. 

\section{APEX/\lowercase{n}FLASH230 Data Reduction and Results} \label{sec:apex-results}

\begin{figure}
\plotone{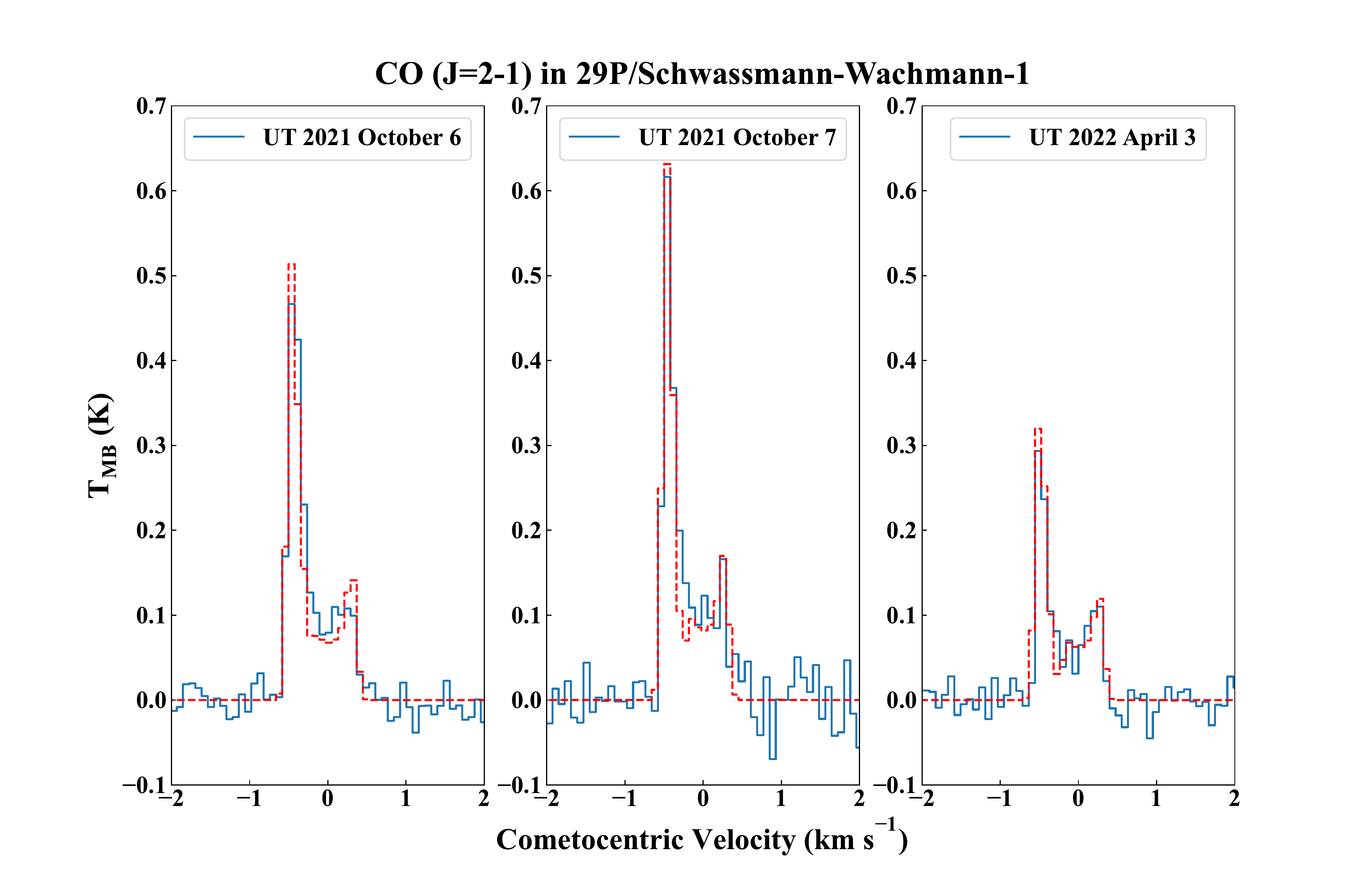}
\caption{Detection of CO (\Ju{}=2--1) in SW1 on 2021 October 6 and 7 and 2022 April 3 with APEX/nFLASH230. The 61 kHz frequency resolution corresponds to a velocity resolution of 0.08 \kms{}. The best-fit radiative transfer model on each date is overlaid in red.
\label{fig:co_apex}}
\end{figure}

We securely detected molecular emission from CO (\Ju{}=2--1) and calculated stringent upper limits for CS, CH$_3$OH, and H$_2$CO (Section~\ref{subsec:trace}). Figure~\ref{fig:co_apex} shows our detections of CO on 2021 October 6 and 7 and 2022 April 3. We modeled molecular line emission using the SUBLIMED three-dimensional radiative transfer code for cometary atmospheres \citep{Cordiner2022}, including a full non-LTE treatment of coma gases, collisions with CO and electrons, and pumping by solar radiation. We used the escape probability formalism \citep{Bockelee1987} to address optical depth effects for the ultra-cold CO emission, along with a time-dependent integration of the energy level population equations. We used explicitly calculated CO-CO collisional rates \citep{Cordiner2022} when modeling CO outgassing. We are unaware of any CH$_3$OH-CO, CS-CO, or H$_2$CO-CO collisional rates available in the literature, so we assumed that they were the same as CH$_3$OH-H$_2$ \citep{Rabli2010}, CS-H$_2$ \citep{Denis-alpizar2018}, and H$_2$CO-H$_2$ \citep{Wiesenfeld2013}, respectively, taken from the LAMDA database \citep{Schoier2005} when calculating 3$\sigma$ upper limits for each molecule. These $X$-H$_2$ collisional rates are the only approximation to collisions of CH$_3$OH, CS, and H$_2$CO with coma neutrals available, and detailed quantum mechanical calculations of $X$-CO collisional rates for these species is beyond the scope of this study. Photodissociation rates for all molecules were adopted from \cite{Huebner2015} with the exception of CS \citep{Boissier2007}, using quiet Sun rates for October and active Sun rates for April.

The CO line profiles are strongly asymmetric with a dominant blue component, ruling out an interpretation of spherically symmetric, isotropic outgassing. We follow the methods of \cite{Cordiner2022} applied to CO-rich comet C/2016 R2 (PanSTARRS), dividing the coma into two outgassing regions, $R_1$ and $R_2$, corresponding to solid angle regions $\Omega_1$ and $\Omega_2$, each with independent molecular production rates ($Q_1$,$Q_2$) and gas expansion velocities ($v_1$,$v_2$). This is consistent with prior interpretation \citep[\eg{}][]{Bockelee2022} that the outgassing consists of (1) a narrow, enhanced, sunward-facing jet (the blueward component, $R_1$) and (2) ambient outgassing from the remainder of the nucleus ($R_2$). The narrow width of the jet and the velocity shift of the line are consistent with CO outflow at or near the subsolar point \citep{Gunnarsson2008} as opposed to uniform outflow. This is the simplest, reasonable model capable of fitting the asymmetric line profiles in SW1.

Region $R_1$ is then defined as the conical region about the subsolar point of the nucleus with half-opening angle $\gamma$ and $R_2$ the remainder of the coma. We assumed a gas kinetic temperature \textit{T}\subs{kin} = 5 K as a middle value between previous work indicating $T\sim$4--6 K \citep{Gunnarsson2008,Paganini2013}. We performed least-squares fits of our radiative transfer models to molecular line profiles, allowing ($Q_1$, $Q_2$, $v_1$, $v_2$, $\gamma$) to vary as free parameters. We determined the distribution of each species in each coma region ($R_1$ or $R_2$) using a Haser formalism \citep{Haser1957}:
\begin{equation}
    n_d(r) = \frac{Q_i}{4 \pi v_i r^2}\frac{\frac{v_i}{\beta_d}}{\frac{v_i}{\beta_d}-L_p}\left[\exp{\left(-\frac{r\beta_d}{v_i}\right)}-\exp{\left(-\frac{r}{L_p}\right)}\right],
\end{equation}
\noindent where $Q_i$ and $v_i$ are the molecular production rate (s$^{-1}$) and gas expansion velocity (km\,s$^{-1}$) in each coma region, \textit{$\beta$}\subs{d} is the molecular photodissociation rate (s$^{-1}$), and \textit{L}\subs{p} is the molecular parent scale length (km). We assumed direct nucleus release (\textit{L}\subs{p} = 0 km) for CO based on previous measurements of SW1 \citep{Bockelee2022}. Figure~\ref{fig:co_apex} shows our extracted CO (\Ju{}=2--1) spectra on each date along with our best-fit radiative transfer models. We calculated integrated intensities of the CO (\Ju{}=2--1) line and production rates $Q_1$ and $Q_2$ in the jet and ambient coma regions, $R_1$ and $R_2$, respectively. $Q_{total}$ is the global production rate. Table~\ref{tab:kinematics} gives our best-fit model parameters. 

\begin{deluxetable*}{cccccccccc}
\tablenum{2}
\tablecaption{Evolution of CO (\Ju{} = 2--1) in 29P/SW1 as Measured by APEX/nFLASH230\label{tab:kinematics}}
\tablewidth{0pt}
\tablehead{
\colhead{Date}  & \colhead{$\int_{R1} T_{MB}\,dv$} & \colhead{$Q_1$} & \colhead{$v_1$} & \colhead{$\int_{R2} T_{MB}\,dv$} & \colhead{$Q_2$} & \colhead{$v_2$} & \colhead{$\gamma$} & \colhead{$Q_{total}$} & \colhead{$Q_1/Q_2$} \\
\colhead{} & \colhead{(K \kms{})} & \colhead{($10^{28}$ s$^{-1}$)} & \colhead{(\kms{})} & \colhead{(K \kms{})} & \colhead{($10^{28}$ s$^{-1}$)} & \colhead{(\kms{})}   & \colhead{($^{\circ}$)} &  \colhead{($10^{28}$ s$^{-1}$)} & \colhead{ }
}
\startdata
2021 Oct 6 & 0.1206 $\pm$ 0.0054 & 3.80 $\pm$ 0.38 & 0.49 $\pm$ 0.01 & 0.0456 $\pm$ 0.0045 & 2.20 $\pm$ 0.23 & 0.34 $\pm$ 0.01 & 55 $\pm$ 2 & 6.00 $\pm$ 0.61 & 1.73 $\pm$ 0.25 \cr
2021 Oct 7 & 0.1316 $\pm$ 0.0089 & 3.44 $\pm$ 0.35 & 0.49 $\pm$ 0.01 & 0.0475 $\pm$ 0.0075 & 2.15 $\pm$ 0.23 & 0.30 $\pm$ 0.02 & 47 $\pm$ 3 & 5.59 $\pm$ 0.57 & 1.60 $\pm$ 0.24 \cr
2022 Apr 3 & 0.0671 $\pm$ 0.0054 & 2.73 $\pm$ 0.28 & 0.53 $\pm$ 0.01 & 0.0334 $\pm$ 0.0046 & 1.78 $\pm$ 0.19 & 0.30 $\pm$ 0.02 & 50 $\pm$ 3 & 4.51 $\pm$ 0.46 & 1.53 $\pm$ 0.22 \cr
\hline
\enddata
\tablecomments{$Q_1$ and $Q_2$ are the molecular production rates and $v_1$ and $v_2$ are the gas expansion velocities in regions $R_1$ and $R_2$, respectively. $\int_{R1} T_{MB}\,dv$ and $\int_{R2} T_{MB}\,dv$ are the integrated intensities of the CO jet ($R_1$) and the ambient CO coma ($R_2$), integrating the line from -0.6 to -0.1 \kms{} and from -0.1 to 0.4 \kms{} for $R_1$ and $R_2$, respectively, and accounting for the line velocity shift. $Q_{total}$ is the global molecular production rate. $Q_1/Q_2$ is the ratio of production rates in regions $R_1$ and $R_2$. An absolute flux calibration uncertainty of 10\% is applied to the molecular production rates.
}
\end{deluxetable*}

\section{IRTF/\lowercase{i}SHELL Data Reduction and Results} \label{sec:ishell-results}
We employed data reduction procedures that have been rigorously tested and are described extensively in the literature \citep{Bonev2005,DiSanti2006,Villanueva2009,Radeva2010,Villanueva2011a,Villanueva2012b,Villanueva2013b,DiSanti2014}, including their application to unique aspects of IRTF/iSHELL spectra \citep{DiSanti2017,Faggi2018,Roth2020}. Each echelle order within an IRTF/iSHELL setting was processed individually as previously described, such that each row corresponded to a unique position along the slit, and each column to a unique wavelength. Spectra were extracted from the processed frames by summing the signal over 15 rows (approximately 2$\farcs$5), seven rows to each side of the nucleus (Figure~\ref{fig:echelle}). We now detail pertinent aspects of the IRTF/iSHELL data reduction unique to our analysis of SW1.

\subsection{Spatial Registration Along the Slit and Extracted Spectra}\label{subsec:spatial_reg}
Although we detected strong molecular emission from CO in SW1, we detected only weak continuum emission, requiring special care in defining the nucleus position along the slit. We accomplished spatial registration using a combination of CO molecular emission and parameters from our bright IR flux standard calibration measurements. We carefully coordinated our flux calibration measurements such that there was no grating change between SW1 exposures and flux standard exposures within an instrumental setting (Table~\ref{tab:obslog}). 

CO emission was present in IRTF/iSHELL M2 setting order 110 (containing the P1, P2, and P3 lines) and order 111 (containing R0 and R1; Table~\ref{tab:ir_co}). We defined the nucleus position as the position of peak CO emission in each order, finding A and B beams separated by 45 rows. For the flux standard spectra in the same orders (110 and 111) we found the rows containing peak stellar continuum emission in each beam were five rows lower than the row containing the peak CO emission in SW1. Thus, for spatial registration in other orders with no detected molecular emission (\ie{} M2 order 106 for OCS) and in the Lp1 setting (sampling C$_2$H$_6$, CH$_3$OH, H$_2$CO, and CH$_4$) we defined the nucleus position by adding five rows to the row containing peak flux standard continuum emission in a given order. 

\begin{figure}
\plotone{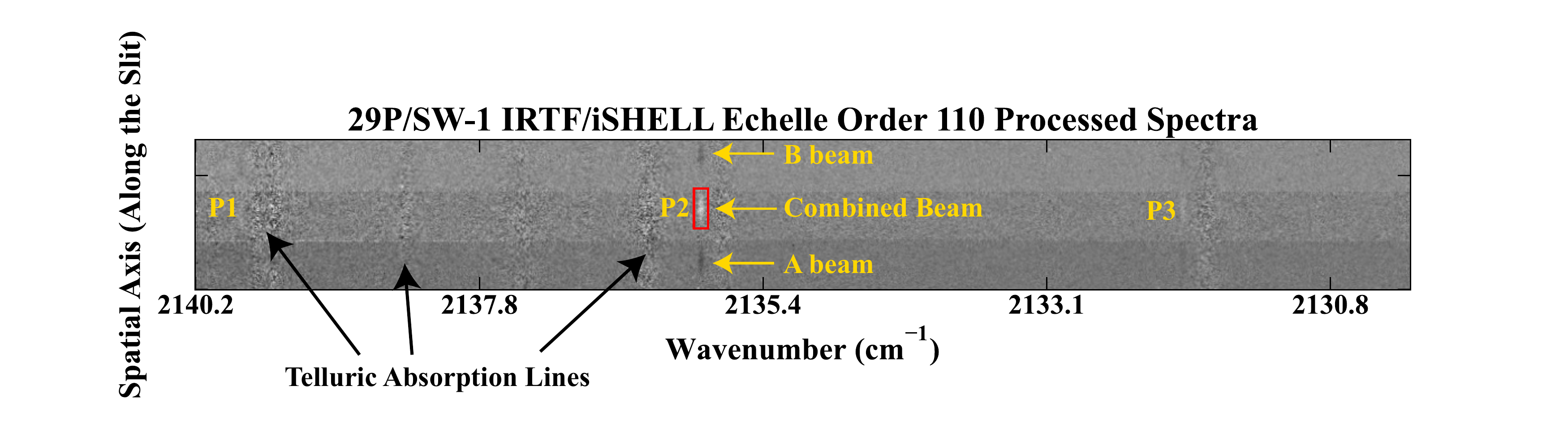}
\caption{Processed IRTF/iSHELL spectra showing detections of CO in SW1 on 2021 October 6 from echelle order 110 (covering 2140--2130 cm$^{-1}$). The positions of the A-beam, B-beam, and combined beam are indicated, individual CO rovibrational transitions are labeled, and the positions of several telluric absorptions are shown. An example nucleus-centered extraction aperture (6-spectral pixels $\times$ 15-spatial pixels, $0\farcs75\times2\farcs5$) is shown for the P2 line in red.
\label{fig:echelle}}
\end{figure}

\begin{figure*}
\gridline{\fig{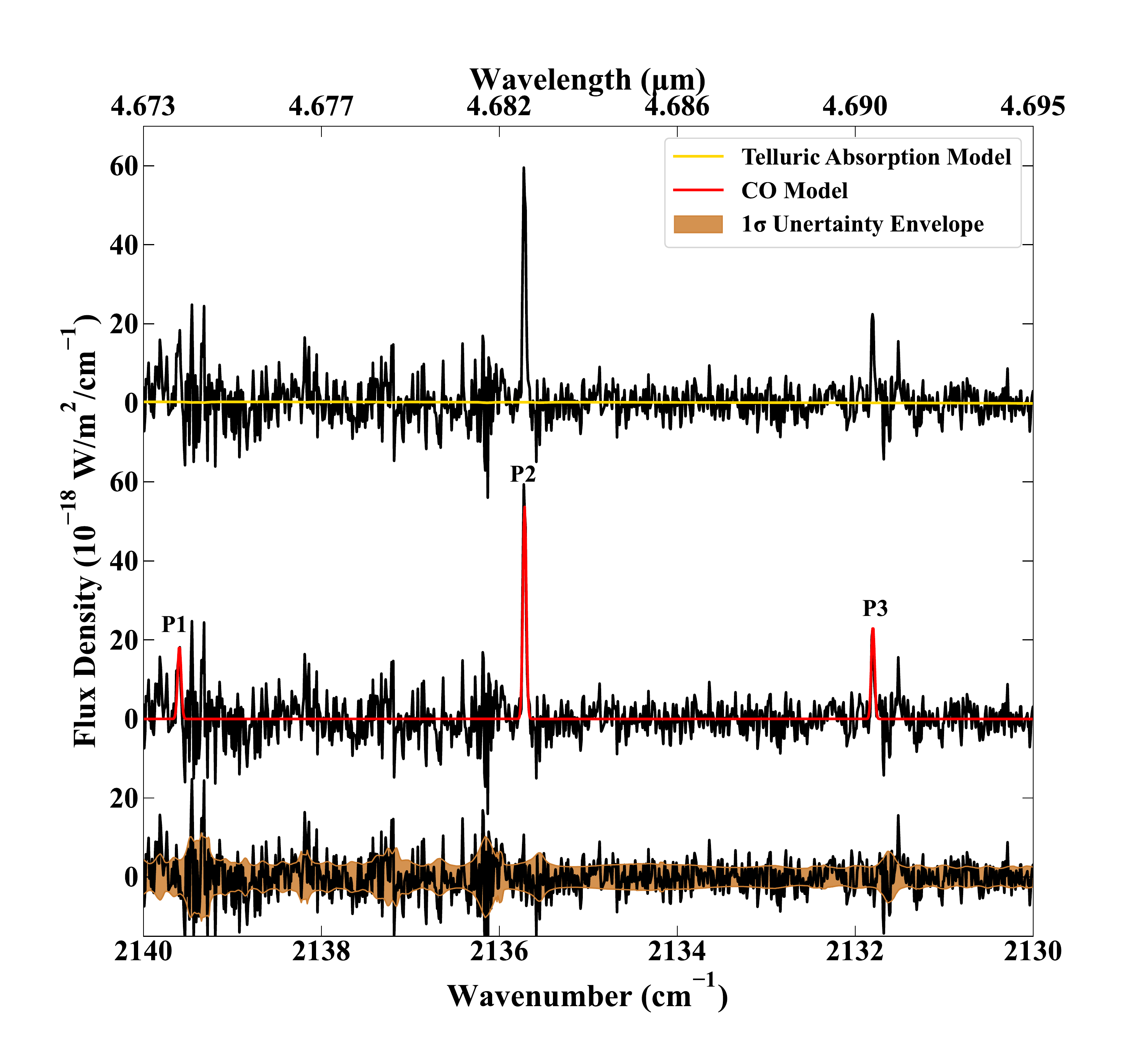}{0.50\textwidth}{(A)}
        \fig{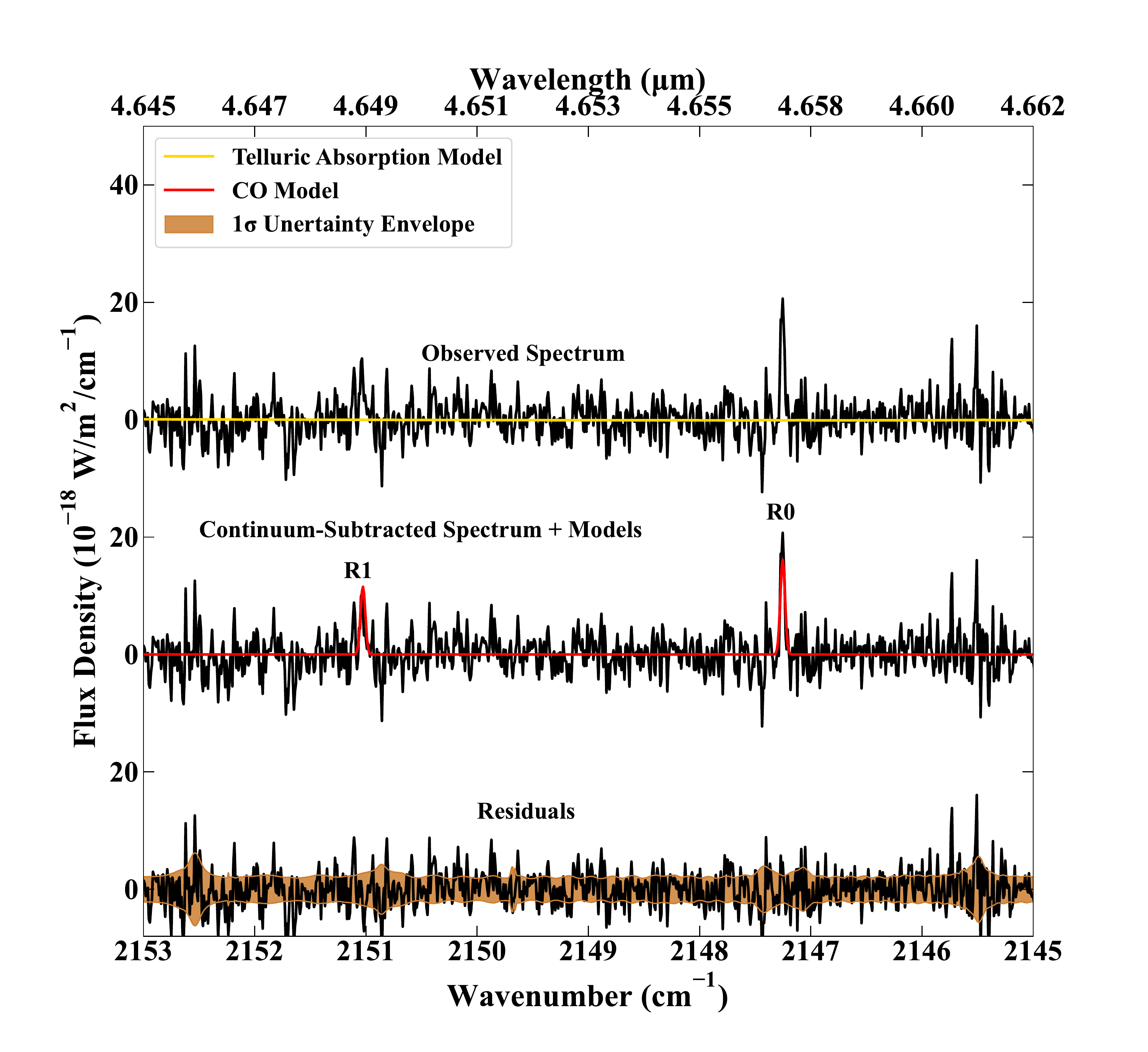}{0.50\textwidth}{(B)}
	}
\caption{\textbf{(A-B).} Extracted IRTF/iSHELL spectra showing detections of CO in SW1 on 2021 October 6 from two separate echelle orders from the M2 setting, (A) order 110 covering 2140--2130 cm$^{-1}$ and (B) order 111 covering 2153 - 2145 cm$^{-1}$. The uppermost spectrum in each panel is the observed spectrum, with the gold trace showing the telluric absorption model (convolved to the instrumental resolution and scaled to the observed continuum level). Telluric features are only weakly visible given the weak continuum emission in SW1. The lower spectrum in each panel is the residual emission spectrum (after subtracting the telluric absorption model) with fluorescence models overlain for CO (red) and the 1$\sigma$ uncertainty envelope shaded in bronze. Individual CO rovibrational transitions are labeled.
\label{fig:co}}
\end{figure*}

We determined contributions from near-infrared continuum emission, telluric extinction, and gaseous emissions in comet spectra as previously described \citep[e.g.,][]{DiSanti2016,DiSanti2017} and illustrate the procedure in Figures~\ref{fig:echelle} and~\ref{fig:co}, showing fully calibrated spectral extracts for CO emission in SW1. Although our spectral setup sampled emissions from H$_2$O, OCS, CH$_4$, C$_2$H$_6$, CH$_3$OH, and H$_2$CO, none of these were detected. Stringent upper limits were derived for OCS, CH$_4$, C$_2$H$_6$, and CH$_3$OH. H$_2$O and H$_2$CO production rates were not meaningfully constrained by our IRTF/iSHELL measurements. We convolved the fully resolved transmittance function to the resolving power of the data ($\sim 4.5 \times 10^4$) and scaled it to the level of the comet continuum. We then subtracted the modeled continuum to isolate cometary emission lines and compared synthetic models of fluorescent emission for each targeted species to the observed line intensities.

\subsection{Molecular Fluorescence Analysis and Rotational Temperature}\label{subsec:fluorescence}
Synthetic models of fluorescent emission for each targeted species were compared to observed line intensities, after correcting each modeled \textit{g}-factor (line intensity) for the monochromatic atmospheric transmittance at its Doppler-shifted wavelength (according to the geocentric velocity of the comet at the time of the observations). The \textit{g}-factors used in synthetic fluorescent emission models in this study were generated with quantum mechanical models developed for CO, OCS, CH$_4$, C$_2$H$_6$, and CH$_3$OH using the NASA Planetary Spectrum Generator \citep[PSG, psg.gsfc.nasa.gov;][]{Villanueva2018}. We applied a special treatment to the CO fluorescence models to account for opacity (Section~\ref{subsec:od}).

A Levenburg-Marquardt nonlinear minimization technique \citep{Villanueva2008} was used to fit fluorescent emission from all species simultaneously in each echelle order, allowing for high precision results, even in spectrally crowded regions containing many spectral lines within a single instrumental resolution element. Production rates for each sampled species were determined from the appropriate fluorescence model at the rotational temperature of each molecule as:
\begin{equation}
    Q = \frac{4\pi \Delta^2 F_i}{g_i \tau(hc\nu) f(x)},
\end{equation}
where \textit{Q} is the molecular production rate (s$^{-1}$), $\Delta$ is the geocentric distance (m), $F_i$ is the flux of line $i$ incident on the terrestrial atmosphere (W\,m$^{2}$), $g_i$ is the \textit{g}-factor (photon\,s$^{-1}$\,molecule$^{-1}$), $\tau$ is the molecular lifetime \citep[s;][]{Huebner2015}, $hc\nu$ is the energy (J) of a photon with wavenumber $\nu$ (cm$^{-1}$), and $f(x)$ is the fraction of the molecules contained within the beam assuming uniform outflow and constant gas expansion velocity. An accurate measure of the gas expansion velocity is critical, particularly in the optically thick case, as $f(x)$ (and therefore $Q$) is sensitive to the gas expansion velocity. Our APEX/nFLASH230 observations enabled a direct measure of the CO expansion velocity, and we adopted the mean value 0.41 \kms{} between the jet and ambient coma regions (Table~\ref{tab:kinematics}) for all of our IR analysis. It is worth noting that the IR formalism assumes spherically symmetric release, as the IR lines are unresolved in velocity space, and the resulting $Q$ is an approximate treatment of the highly asymmetric outgassing and $Q_{total}$ measured with APEX/nFLASH230. However, the $Q$-curve formalism (Sections~\ref{subsubsec:growth} and~\ref{subsubsec:pump}) provides at least a first-order treatment to asymmetric outgassing by averaging emission intensity on both sides of the nucleus when deriving a Growth Factor.

Rotational temperatures (\textit{T}\subs{rot}) were determined using correlation and excitation analyses as described in \cite{Bonev2005,Bonev2008,DiSanti2006,Villanueva2008}. In the case of SW1, determination of the CO rotational temperature was hampered by optical depth effects: as detailed in Section~\ref{subsec:od}, correcting for the opacity requires either spectral extracts taken far from the nucleus or from small apertures, resulting in a lower signal-to-noise ratio than the usual case of a 15 pixel (2.5$\arcsec$) nucleus-centered extract used for comets with optically thin comae. Within the limitations of our data, we find that \trot{} = 3--5 K is consistent with the relative intensities of the CO lines, consistent with previous near-infrared observations \citep{Paganini2013}, and we assume \trot{} = 4 K in all instances and for all molecules in analysis of our IRTF/iSHELL spectra.

\subsection{Treatment of Optical Depth}\label{subsec:od}
The ultra-cold CO emission in SW1 was likely optically thick along lines of sight passing close to the nucleus, similar to that observed in other CO-rich comets \citep[\eg{} C/2016 R2 (PanSTARRS), C/1995 O1 (Hale-Bopp);][]{McKay2019,DiSanti2001}, which can affect the derived molecular production rates. \cite{Bockelee2010} and \cite{DiSanti2001} demonstrated the importance of treating opacity effects for CO emission in these comets.

We corrected for optical depth using three complementary approaches, including two developed for similarly distant, CO-rich comets: (1) The $Q$-curve formalism described in \cite{Bonev2017} to analyze optically thick CO in comet C/2003 W6 (Christensen); (2) Correction for opacity in the solar pump, developed to analyze CO emission in C/1995 O1 (Hale-Bopp) and C/1996 B2 (Hyakutake) \citep{DiSanti2001,DiSanti2003}; and (3) Implementation of a new treatment for optical depth correction in the NASA PSG \citep{Villanueva2018,Villanueva2022} applied to extracted column densities along the slit. Both approaches (1) and (2) have subsequently been applied to the case of optically thick CO emission in peculiar, ultra CO-rich comet C/2016 R2 \citep[PanSTARRS;][]{McKay2019} and used CO models generated by the PSG with optically thin $g$-factors, whereas approach (3) used CO models generated by the PSG with $g$-factors corrected for opacity. In all instances the CO models were adjusted for the heliocentric velocity of the comet at the time of our observations (0.49 \kms{}) to correct for the Swings effect. We demonstrate consistency among all approaches and the robustness of our results. 

\subsubsection{$Q$-Curve Analysis}\label{subsubsec:growth}
\begin{figure*}
\gridline{\fig{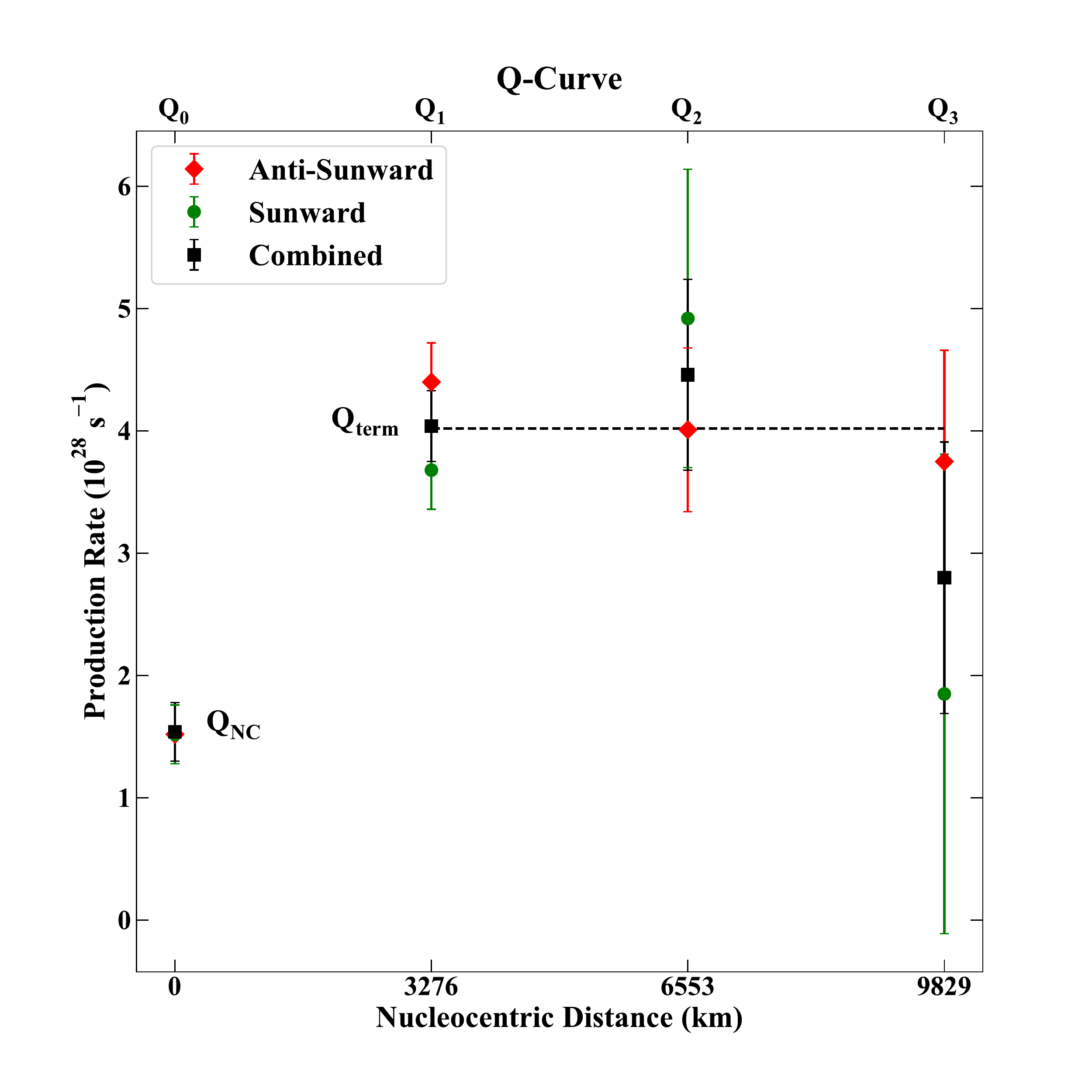}{0.50\textwidth}{(A)}
          \fig{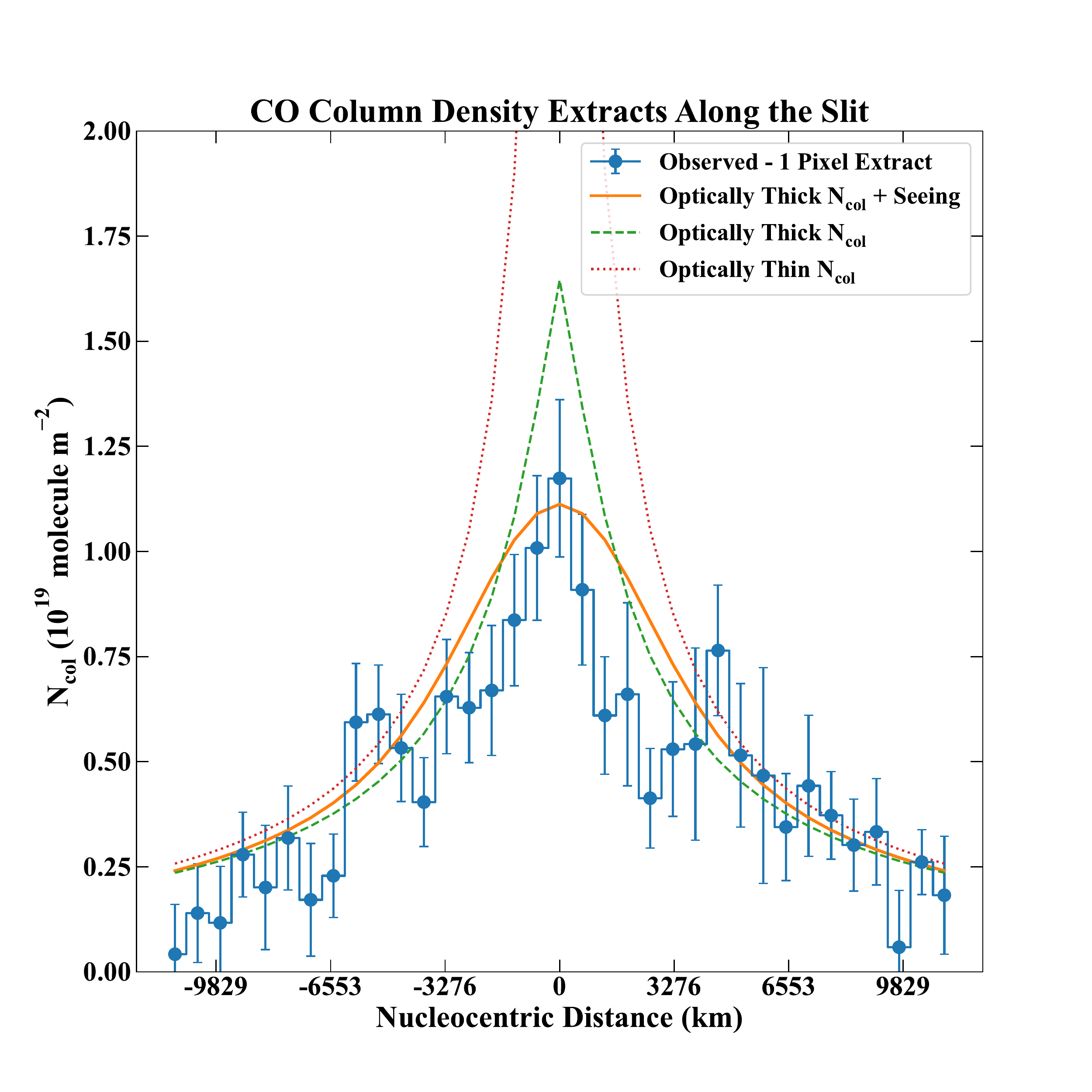}{0.5\textwidth}{(B)}
	}
\caption{\textbf{Left.} $Q$-curve constructed using the curve-of-growth approach and showing retrieved symmetric \qco{} vs. nucleocentric distance in SW1. Nucleus-centered ($Q_{NC}$) and terminal ($Q_{term}$) value are indicated in the figure with 1$\sigma$ uncertainties. \textbf{Right.} Extracts of CO column density ($N_{col}$) in 1 pixel (0\farcs167) increments along the slit along with 1$\sigma$ uncertainties. Also shown are models for optically thick $N_{col}$ (including after convolution with the seeing) and optically thin $N_{col}$. The optically thick models were generated with opacity corrections using the NASA PSG (Appendix~\ref{sec:odpsg}). An acceptable fit is found for \qco{} = $4.8\times10^{28}$ s$^{-1}$ and \trot{} = 4 K, assuming an expansion velocity $v$ = 0.41 \kms{} and 1$\farcs$1 seeing. Note that the off-nucleus regions in which optically thin and optically thick models are in agreement are consistent with the terminal region in the $Q$-curves from the curve-of-growth and in Figure~\ref{fig:co_qcurve}.
\label{fig:true_qcurve}}
\end{figure*}

The $Q$-curve formalism described in \cite{Bonev2017} obtains $Q$ not based on the nucleus-centered extracts (where line fluxes are affected by optical depth), but from spectra offset sufficiently from the nucleus that optically thin conditions are reached. The common application of the $Q$-curve formalism involves summing the fluxes of all emission lines to increase the signal-to-noise ratio, and assuming a constant $g$-factor along the slit. This approach provides an excellent approximation in the optically thin case. However, here we construct a ``true'' $Q$-curve by first extracting spectra at successive field-of-view intervals along the slit. The $Q$-curve is then symmetrized by averaging extracts taken on each side of the nucleus. Importantly, production rates are obtained through line-by-line analysis (\ie{} without summing the fluxes). The resulting $Q$-curve is shown in Figure~\ref{fig:true_qcurve}.

The symmetric \textit{Q}-values increase with nucleocentric distance owing primarily to atmospheric seeing and opacity, which suppress signal along lines of sight passing close to the nucleus due to the use of a narrow slit, until reaching a terminal value. The ratio between emission intensity at the terminal position to that at the nucleus-centered position is taken to be the multiplicative Growth Factor \citep[\eg{}][]{Villanueva2011a,DiSanti2016}. This multiplicative Growth Factor is applied to nucleus-centered spectral extracts when calculating global production rates for each molecule. 

We calculated \qco{} for 5 pixel (0\farcs83) extracts to either side of and equidistant from the nucleus in successive increments using CO models generated by the PSG with optically thin $g$-factors. Our nucleus-centered region is then defined as $Q_0$, and our ``terminal'' region is defined as the weighted average of $Q_1$, $Q_2$, and $Q_3$ (Figure~\ref{fig:true_qcurve}). \qco{} in the nucleus-centered region (affected by optical depth and slit losses) is $(1.52\pm0.24)\times10^{28}$ s$^{-1}$, \qco{} in the terminal, optically thin region is $(4.02\pm0.26)\times10^{28}$ s$^{-1}$, and the Growth Factor, GF = $Q_{terminal}$/$Q_{NC}$ is 2.64 $\pm$ 0.46.

\subsubsection{Correction for Opacity in the Solar Pump}\label{subsubsec:pump}
\begin{figure}
\plotone{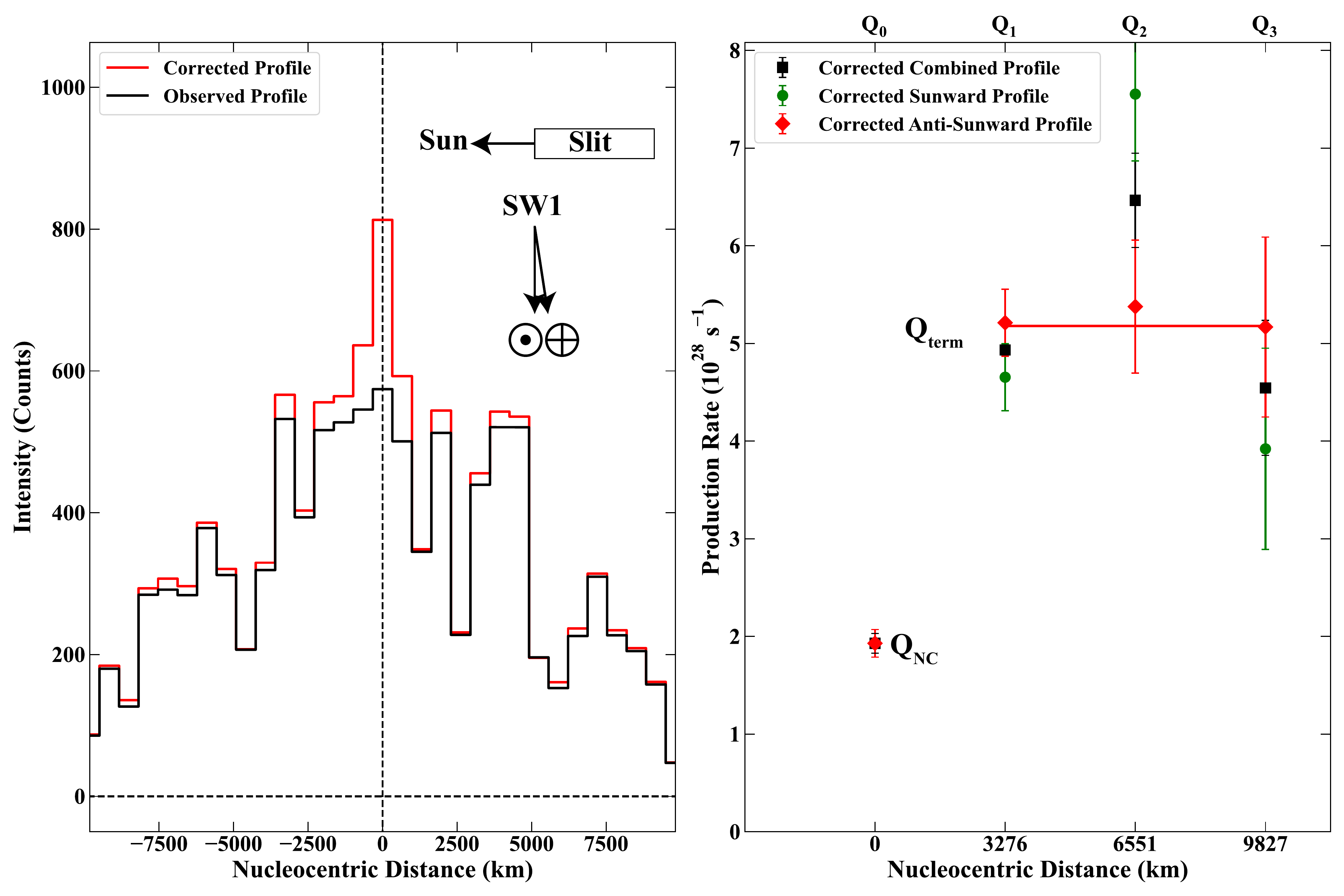}
\caption{\textbf{Left.} Spatial profiles of emission for CO (summed over the P1, P2, and P3 lines) in SW1 measured with IRTF/iSHELL. The observed spatial profile (black) and spatial profile corrected for opacity in the solar pump (red) are shown. the slit was oriented along the projected Sun-comet line (PA 259$^{\circ}$). The solar phase angle of 8.7$^{\circ}$ is indicated. \textbf{Right.} $Q$-curve calculated using the opacity corrected spatial profile. The terminal ($Q_1$--$Q_3$) and nucleus-centered ($Q_0$) values are indicated.
\label{fig:co_qcurve}}
\end{figure}

In our second case, we adopted the methodology in the Appendix of \cite{DiSanti2001}, calculating the ``critical distance'' from the nucleus for which each CO transition reaches unit opacity and correcting the solar pump assuming uniform gas outflow at constant speed. These corrections were applied to spatial profiles of emission intensity for each CO transition separately, enabling the calculation of a ``corrected'' \textit{Q}-curve and multiplicative (GF) from the summed profiles (Figure~\ref{fig:co_qcurve}). Our corrected GF, 2.68 $\pm$ 0.23, is consistent with GF's measured in observations of optically thin emission in other comets with IRTF/iSHELL \citep[\eg{}][]{Roth2021b} and with that derived in the previous section. Applying the corrected GF to a classical 15-pixel nucleus-centered extract using CO models with optically thin $g$-factors, we derive \qco{} = $(5.18\pm0.69)\times10^{28}$ s$^{-1}$.

\subsubsection{Correction for Opacity in the NASA Planetary Spectrum Generator}\label{subsubsec:column}
As a further validation of the production rates obtained in Sections~\ref{subsubsec:growth} and~\ref{subsubsec:pump}, we developed and implemented an approximate treatment for optical depth in the NASA PSG. Our formalism is described in Appendix~\ref{sec:odpsg}. The corrections must be applied over a relatively small field of  view to prevent beam dilution of the column density. We calculated the CO column density in sliding 1-pixel ($0\farcs167$) extracts along the slit. We then varied \qco{} until the modeled optically thick column density (convolved with the seeing, which we approximate to be $\sim$1$\farcs$1) provided a reasonable fit to the observed column density profiles. Figure~\ref{fig:true_qcurve} shows our results, with an acceptable fit for \qco{} = $4.8\times10^{28}$ s$^{-1}$, where we estimate an uncertainty on the order of 10\%. Table~\ref{tab:ir_co} provides our measured CO line fluxes in a nucleus-centered, 15-pixel extract along with our optically thin and corrected $g$-factors at the nucleus position.

\begin{deluxetable*}{ccccc}
\tablenum{3}
\tablecaption{IRTF/iSHELL CO Line Fluxes and $g$-factors\label{tab:ir_co}}
\tablewidth{0pt}
\tablehead{
\colhead{ID} & \colhead{$\nu$} & \colhead{Flux} & \colhead{$g$\subs{thin}\sups{a}} & \colhead{$g$\subs{thick}\sups{b}} \\
\colhead{} & \colhead{(cm$^{-1}$)} & \colhead{($10^{-19}$ W\,m$^{-2}$)} & \colhead{(s$^{-1}$)} & \colhead{(s$^{-1}$)}
}
\startdata
P3 & 2131.63 & 11.5 $\pm$ 1.5 & $3.55\times10^{-5}$ & $1.13\times10^{-5}$ \\
P2 & 2135.54 & 33.4 $\pm$ 1.6 & $8.42\times10^{-5}$ & $1.44\times10^{-5}$ \\
P1 & 2139.42 & 10.3 $\pm$ 1.9 & $3.06\times10^{-5}$ & $1.59\times10^{-5}$ \\
R0 & 2147.08 & 9.68 $\pm$ 0.68 & $4.28\times10^{-5}$ & $7.31\times10^{-6}$ \\
R1 & 2150.86 & 5.09 $\pm$ 0.89 & $2.44\times10^{-5}$ & $7.78\times10^{-6}$ \\
\hline
\enddata
\tablecomments{\sups{a}Optically thin $g$-factor. \sups{b}Optically thick $g$-factor at the nucleus position corrected using the formalism described in Appendix~\ref{sec:odpsg}.
}
\end{deluxetable*}

\section{Discussion}\label{sec:discussion}
Our multi-wavelength observations of SW1 during its exceptional 2021 September--October outburst, combined with follow-up observations in 2022 April, enabled a comparison between outburst and quiescent activity in SW1 itself as well as a comparison of its relatively primitive Kuiper disk material against that contained in JFCs and OCCs. We discuss the outburst in the context of longstanding observing programs targeting SW1 and place our compositional measurements into context with the comet population.

\subsection{Optical Context and Comparison with Previous Radio Observations of SW1}\label{subsec:outburst}
\cite{Bockelee2022} derived a relationship between $m_R(1,\rh{},0)$ (SW1's apparent $R$ magnitude corrected to $\Delta$ = 1 au and $\phi$ = 0$^{\circ}$), its expected total \qco{}, and the portion of \qco{} attributable to an outburst. \cite{Miles2021} reported apparent $R$ magnitudes $m_R(1,1,0)$ within a 10\arcsec diameter aperture corrected to \rh{} = $\Delta$ = 1 au and $\phi$ = 0$^{\circ}$ daily throughout September, October, and April. We converted these to $m_R(1,\rh{},0)$ as

\begin{equation}
m_R(1,\rh{},0) = m_R(1,1,0) + 5\log\rh{}
\end{equation}

and used the formalism in \cite{Bockelee2022} to calculate total \qco{} and the outburst portion during our observations. 

In order to compare our results to \cite{Bockelee2022} we performed radiative transfer calculations using the same kinematical parameters, namely fixing $v_1$ = 0.50 \kms{}, $v_2$ = 0.30 \kms{}, $\gamma$ = 45$^{\circ}$, and assuming that $Q_1$ (the CO jet) accounts for 60\% of the total \qco{}. Table~\ref{tab:compare} provides a comparison of our nominal best-fit radiative transfer models against the results when assuming the kinematics in \cite{Bockelee2022}. Notably, our total \qco{} is considerably lower when adopting these kinematics on October 6 and April 3 when our $\gamma$ is larger than 45$^{\circ}$, whereas values for both formalisms are in good agreement for October 7, when our $\gamma$ is consistent with 45$^{\circ}$. This demonstrates that the differences in our calculated \qco{} arise from our treatment of the kinematics. 

\begin{figure*}
\gridline{\fig{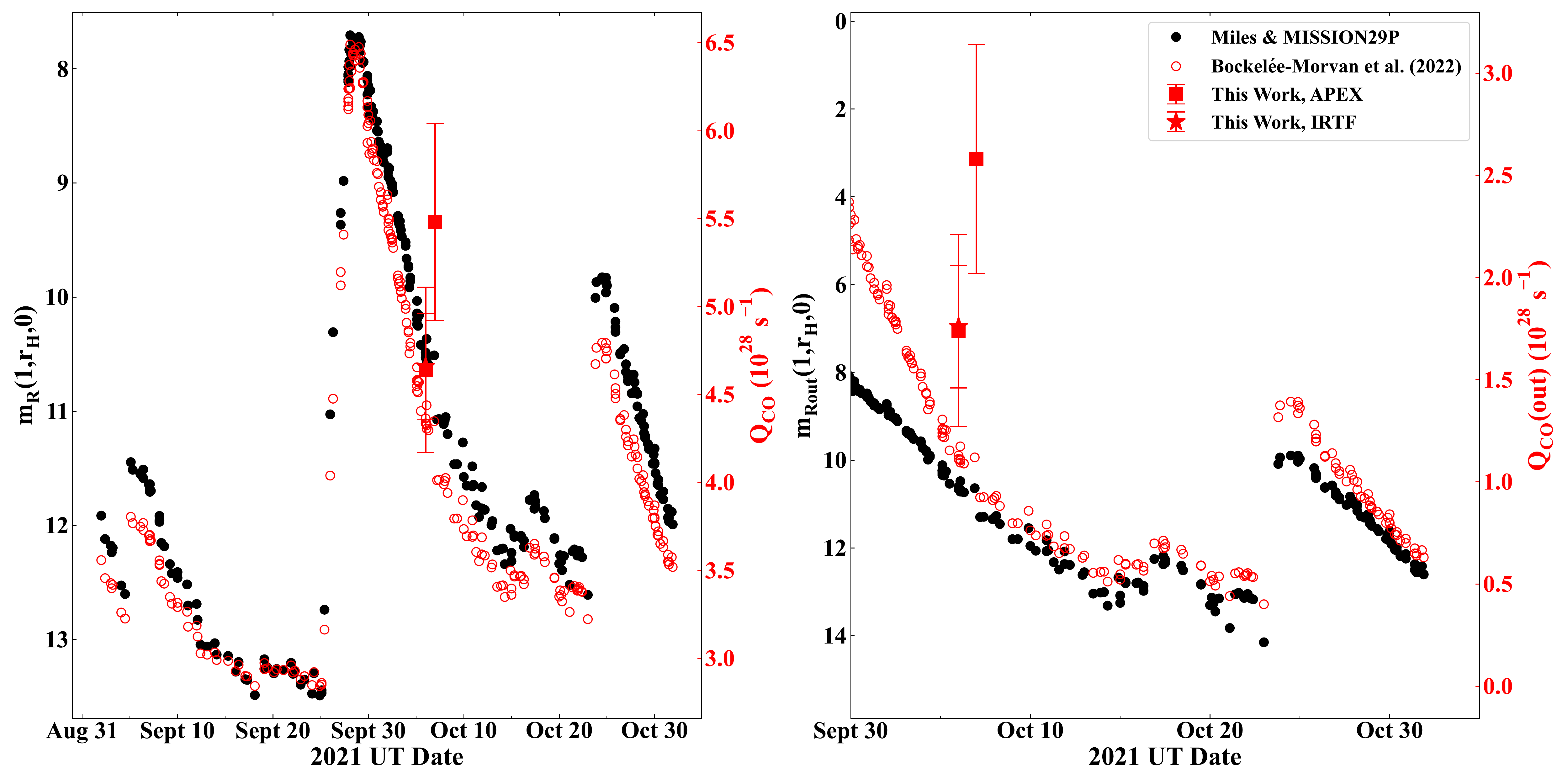}{\textwidth}{(A)}
	}
\gridline{\fig{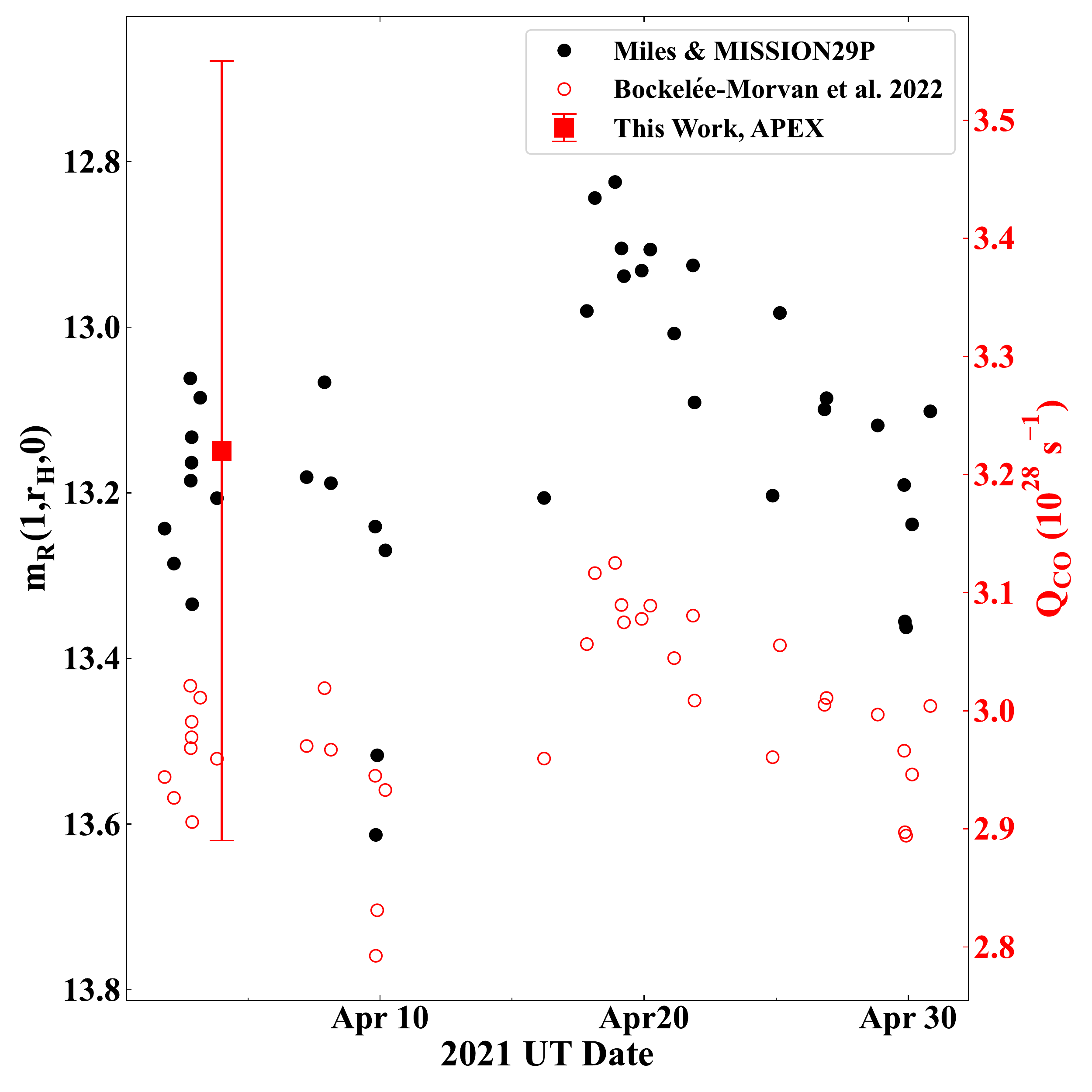}{0.50\textwidth}{(B)}
	}
\caption{\textbf{(A). Left.} Scaled $R$ total magnitudes \citep{Miles2021}, empirically predicted total \qco{} \citep[Section~\ref{subsec:outburst};][]{Bockelee2022}, and measured \qco{} by APEX/nFLASH230 and IRTF/iSHELL (this work, Tables~\ref{tab:compare} and~\ref{tab:comp}) during SW1's September -- October outburst. \textbf{Right.} Portion of the $R$ magnitudes, predicted \qco{}, and measured \qco{} (this work) attributable to outburst \citep[above quiescent activity, defined as \qco{} = $2.9\times10^{28}$ s$^{-1}$;][]{Wierzchos2020} following \cite{Bockelee2022}. \textbf{(B).} Same as upper panel for the April observations.
\label{fig:mag1}}
\end{figure*}

For consistency in comparison against the literature, we adopt \qco{} calculated following \cite{Bockelee2022} for all further discussion. Figure~\ref{fig:mag1} compares our APEX/nFLASH230 total \qco{}, along with \qco{} from our IRTF/iSHELL results, against the values predicted using the relationship between \qco{} and $m_R(1,\rh{},0)$. Our total \qco{} from IRTF/iSHELL and APEX/nFLASH230 are in formal agreement with the predictions for October 6 and April 3, but our APEX/nFLASH230 value on October 7 is still higher than the predicted value. This higher than expected \qco{} may be owing to the exceptional nature of the outburst. \cite{Bockelee2022} measured \qco{} ranging from (3.0 -- 3.3)$\times10^{28}$ s$^{-1}$ on 2021 November 13--15, which is consistent with our \qco{} in April and suggests that SW1 had returned to quiescent activity by November.

\begin{deluxetable*}{cccc|ccc}
\tablenum{4}
\tablecaption{Comparison of Radiative Transfer Models for CO (\Ju{} = 2--1) in 29P/SW1\label{tab:compare}}
\tablewidth{0pt}
\tablehead{
\colhead{Date} & \colhead{$\gamma$} & \colhead{$Q_1$} & \colhead{$Q_{total}$} & \colhead{$\gamma$} & \colhead{$Q_1$} & \colhead{$Q_{total}$} \\
\colhead{} & \colhead{($^{\circ}$)} & \colhead{($10^{28}$ s$^{-1}$)} & \colhead{($10^{28}$ s$^{-1}$)} & \colhead{}  & \colhead{($10^{28}$ s$^{-1}$)} & \colhead{($10^{28}$ s$^{-1}$)} 
}
\startdata
\hline
2021 Oct 6 & 55 $\pm$ 2 & 3.80 $\pm$ 0.38 & 6.00 $\pm$ 0.61 & 45 & 2.78 $\pm$ 0.28 & 4.64 $\pm$ 0.47 \\
2021 Oct 7 & 47 $\pm$ 3 & 3.44 $\pm$ 0.35 & 5.59 $\pm$ 0.58 & 45 & 3.29 $\pm$ 0.33 & 5.48 $\pm$ 0.56  \\
2022 Apr 3 & 50 $\pm$ 3 & 2.73 $\pm$ 0.28 & 4.51 $\pm$ 0.46 & 45 & 1.93 $\pm$ 0.20 & 3.22 $\pm$ 0.33  \\
\enddata
\tablecomments{\textbf{Left.} $Q_1$ and $Q_{total}$ for the nominal best-fit radiative transfer model with indicated jet half-opening angle $\gamma$ and $v_1$, $v_2$, and $Q_1$/$Q_2$ as in Table~\ref{tab:kinematics}. \textbf{Right.} $Q_1$ and $Q_{total}$ fixing $\gamma$ = 45$^{\circ}$, $v_1$ = 50 \kms{}, $v_2$ = 30 \kms{}, and assuming that $Q_1$ represents 60\% of $Q_{total}$ as in \cite{Bockelee2022}.
}
\end{deluxetable*}

\subsection{Comparison of Trace Species Abundances and Comets Measured}\label{subsec:trace}
The high activity during SW1's outburst coupled with the sensitivity of APEX/nFLASH230 and IRTF/iSHELL enabled us to obtain sensitive 3$\sigma$ upper limits on the production of OCS, CS, CH$_3$OH, H$_2$CO, CH$_4$, and C$_2$H$_6$ relative to CO. Our upper limits for CS/CO ($<$ 0.8--4\%) and OCS/CO ($<$ 4.8\%) are the first reported in the literature for SW1. Our upper limits for CH$_3$OH/CO ($<$ 1.8--11\%), C$_2$H$_6$/CO ($<$ 1.96\%) and CH$_4$/CO ($<$ 0.98\%) are significantly more stringent than in previous work \citep{Paganini2013}. 

\begin{deluxetable*}{ccccccc}
\tablenum{5}
\tablecaption{Molecular Composition of SW1 as Measured by IRTF/iSHELL\label{tab:comp}}
\tablewidth{0pt}
\tablehead{
\colhead{Setting} & \colhead{Species} & \colhead{\textit{T}\subs{rot}\sups{a}} & \colhead{GF\sups{b}} & \colhead{\textit{Q$_{total}$}\sups{c}} & \colhead{$Q_x/Q_{CO}$} & \colhead{Range in Comets}\\
\colhead{} & \colhead{ } & \colhead{(K)} & \colhead{ } & \colhead{(10$^{26}$ s\sups{-1})}  & \colhead{(\%)} & \colhead{(\%)}
}
\startdata
\multicolumn{7}{c}{2021 October 6, \rh{} = 5.91 au, $\Delta$ = 5.41 au} \\
\hline
M2 & CO & (4) & 2.64 $\pm$ 0.34 & 466 $\pm$ 30 & 100  & $\cdot \cdot \cdot$ \\
   & OCS & (4) & (2.64) & $<$ 22 (3$\sigma$) & $<$ 4.81 (3$\sigma$) & 1.5--14 \\
\hline
Lp1 & CH$_4$ & (4) & (2.64) & $<$ 4.6 (3$\sigma$) & $<$ 0.98 (3$\sigma$) & 4.6--164 \\
    & C$_2$H$_6$ & (4) & (2.64) & $<$ 9.1 (3$\sigma$) & $<$1.96 (3$\sigma$) & 2.3--98 \\
    & CH$_3$OH & (4) & (2.64) & $<$ 21 (3$\sigma$) & $<$ 4.4 (3$\sigma$) & 10--500 \\
\hline
\enddata
\tablecomments{\sups{a}Rotational temperature.  \sups{b}Growth factor assumed from Section~\ref{subsubsec:growth}. \sups{c}Production rate. We calculated an average \qco{} based on each of the three methods used to address optical depth (Section~\ref{subsec:od}). \sups{d}Mixing ratio with respect to CO (CO = 100). Assumed values are in parentheses.
}
\end{deluxetable*}

\cite{Bockelee2022} reported H$_2$O and HCN production in SW1 consistent with spherically symmetric sublimation from icy grains at \textit{T}\subs{kin} = 100 K and \lp{} = 10,000 km. We calculated 3$\sigma$ upper limits for CH$_3$OH, H$_2$CO, and CS production in two cases: (1) Assuming direct nucleus release with the same temperature as CO, and (2) Assuming production from icy grains with the same \lp{} and \textit{T}\subs{kin} reported for H$_2$O and HCN and assuming the average expansion velocity derived for CO on each date. Upper limits for CH$_4$ and C$_2$H$_6$ constrained by IRTF/iSHELL were calculated assuming spherically symmetric, direct nucleus release at \trot{} = 4 K (Table~\ref{tab:comp}). Table~\ref{tab:apex_comp} details our full compositional results for APEX/nFLASH230. 

\begin{deluxetable*}{cccccccc}
\tablenum{6}
\tablecaption{Molecular Composition in 29P/SW1 as Measured by APEX/nFLASH230\label{tab:apex_comp}}
\tablewidth{0pt}
\tablehead{
\colhead{Transition} & \colhead{$E_u$} & \colhead{$\nu$} & \colhead{$\int T_{MB}\,dv$}  & \colhead{$Q_{parent}$} & \colhead{$Q_{Lp=10,000 km}$} & \colhead{$Q_x/Q_{CO}$} & \colhead{Range in Comets} \\
\colhead{} & \colhead{(K)} & \colhead{(GHz)} & \colhead{(K \kms{})} & \colhead{($10^{26}$ s$^{-1}$)} & \colhead{($10^{26}$ s$^{-1}$)} & \colhead{(\%)}  & \colhead{(\%)} 
}
\startdata
\multicolumn{7}{c}{2021 October 6, \rh{} = 5.91 au, $\Delta$ = 5.41 au, \qco{} = $4.64\times10^{28}$\,s$^{-1}$} \cr
\hline
CH$_3$OH ($J_K$=$5_0$--$4_0$ $A^+$) & 34.8 & 241.791 & $<$ 0.046 (3$\sigma$) &  $<$ 121 (3$\sigma$) & $<$ 4.8 (3$\sigma$) & $<$ 4.8--26 (3$\sigma$) & 10--500 \\
H$_2$CO ($J_{Ka,Kc}$=3$_{0,3}$--2$_{0,2}$) & 21.0 & 218.222 & $<$ 0.032 (3$\sigma$) &  $<$ 6.80 (3$\sigma$) & $<$ 7.42 (3$\sigma$) & $<$ 1.4--1.6 (3$\sigma$) & 1--81  \\ 
CS (\Ju{}=5--4) & 35.3 & 244.935 & $<$ 0.048 (3$\sigma$) &  $<$ 83 (3$\sigma$) &  $<$ 9.7 (3$\sigma$) & $<$ 2.1--18 (3$\sigma$) & 0.02--5.5 \\
\hline
\multicolumn{7}{c}{2021 October 7, \rh{} = 5.91 au, $\Delta$ = 5.39 au, \qco{} = $5.48\times10^{28}$\,s$^{-1}$} \cr
\hline
CH$_3$OH ($J_K$=$5_0$--$4_0$ $A^+$) & 34.8 & 241.791 & $<$ 0.026 (3$\sigma$) &  $<$ 59 (3$\sigma$) & $<$ 10 (3$\sigma$) & $<$ 1.8--10.8 (3$\sigma$) & 10--500 \\
H$_2$CO ($J_{Ka,Kc}$=3$_{1,2}$--2$_{1,1}$) & 33.4 & 225.697 & $<$ 0.028 (3$\sigma$) & $<$ 11 (3$\sigma$) & $<$ 8.8 (3$\sigma$) & $<$ 1.6--2.0 (3$\sigma$) & 1--81  \\ 
CS (\Ju{}=5--4) & 35.3 & 244.935 & $<$ 0.022 (3$\sigma$) & $<$ 22 (3$\sigma$) &  $<$ 41 (3$\sigma$) & $<$ 0.7--4.0 (3$\sigma$) & $<$ 0.02--5.5 \\
\hline
\multicolumn{7}{c}{2022 April 3, \rh{} = 5.97 au, $\Delta$ = 6.43 au, \qco{} = $3.22\times10^{28}$\,s$^{-1}$} \cr
\hline
CH$_3$OH ($J_K$=$5_0$--$4_0$ $A^+$) & 34.8 & 241.791 & $<$ 0.017 (3$\sigma$) & $<$ 50 (3$\sigma$) & $<$ 9.6 (3$\sigma$) & $<$ 3--15 (3$\sigma$) & 10--500 \\
H$_2$CO ($J_{Ka,Kc}$=3$_{1,2}$--2$_{1,1}$) & 33.4 & 225.697 & $<$ 0.016 (3$\sigma$) & $<$ 15 (3$\sigma$) & $<$ 9.9 (3$\sigma$) & $<$ 3.1--3.3 (3$\sigma$) & 1--81  \\ 
CS (\Ju{}=5--4) & 35.3 & 244.935 & $<$ 0.014 (3$\sigma$) & $<$ 49 (3$\sigma$) & $<$ 5.8 (3$\sigma$) & $<$ 1.8--11 (3$\sigma$) & $<$ 0.02--5.5 \\
\enddata
\tablecomments{$E_u$, $\nu$ and $\int T_{MB}\,dv$ are the upper state energy, frequency, and integrated intensity (from -0.6 to 0.4 \kms{}) of each molecular transition. $Q_{parent}$ and $Q_{Lp=10,000 km}$ are production rates calculated assuming direct nucleus release and icy grain release at \lp{} = 10,000 km, respectively. $Q_x/Q_{CO}$ is the abundance with respect to CO given as a range between the direct release and icy grain values, with \qco{} taken as the value calculated using kinematics from \cite{Bockelee2022} (Table~\ref{tab:compare}). The range of abundances with respect to CO in measured comets is provided \citep{Bockelee2017,DelloRusso2016a}.
}
\end{deluxetable*}

Our stringent upper limits on the hypervolatile (CO, CH$_4$, C$_2$H$_6$) and oxygen-bearing species (CH$_3$OH and H$_2$CO) composition in SW1 enable us to place the composition of primitive Kuiper disk material into context with measured comets from the major dynamical classes, JFCs and OCCs. A significant caveat is the difference in \rh{}: the majority of measured comets are studied at smaller \rh{} in H$_2$O-dominated comae. Nevertheless, the comparison is worthwhile. 

\begin{figure*}
\gridline{\fig{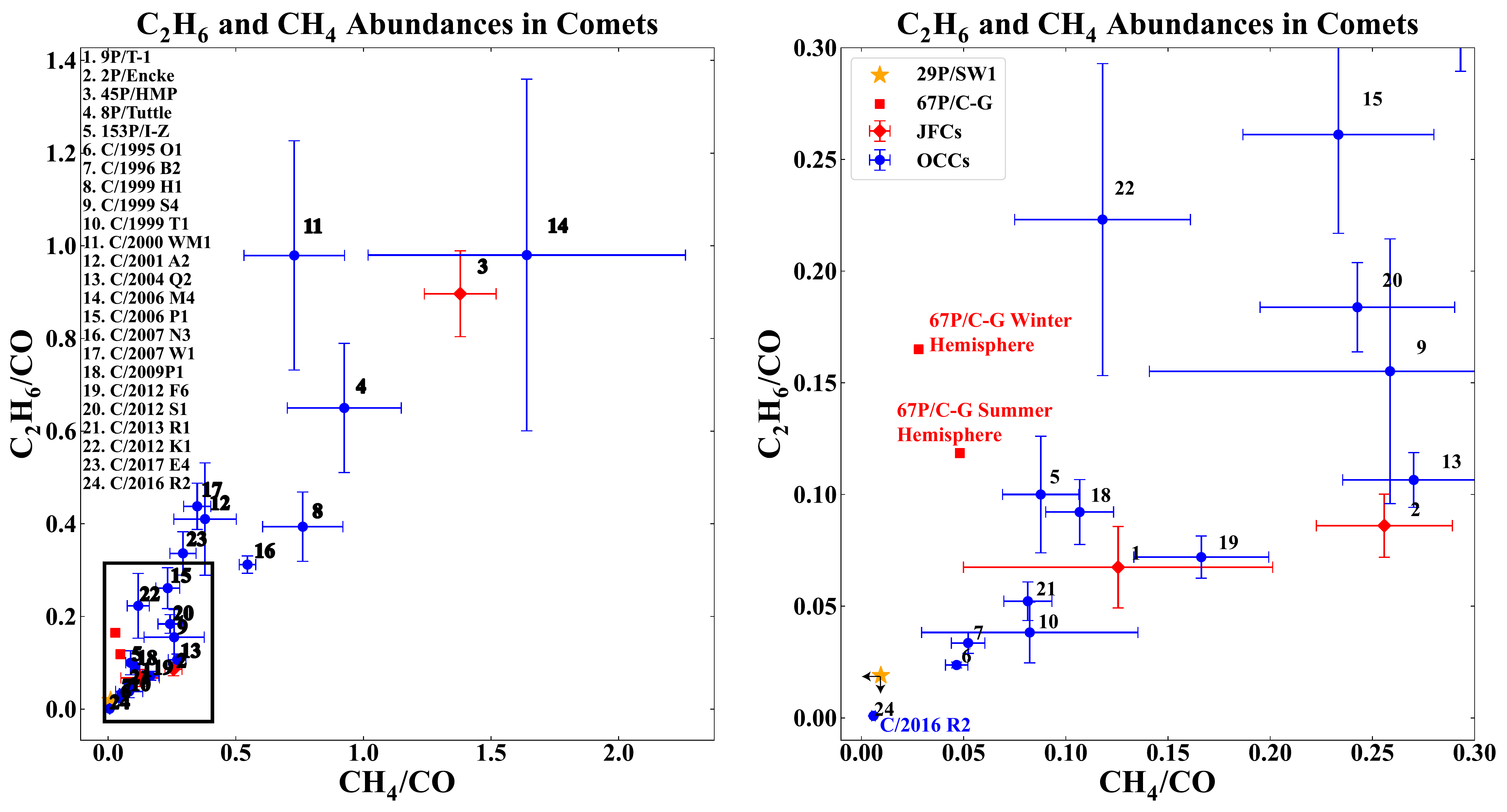}{\textwidth}{(A)}
	}
\gridline{\fig{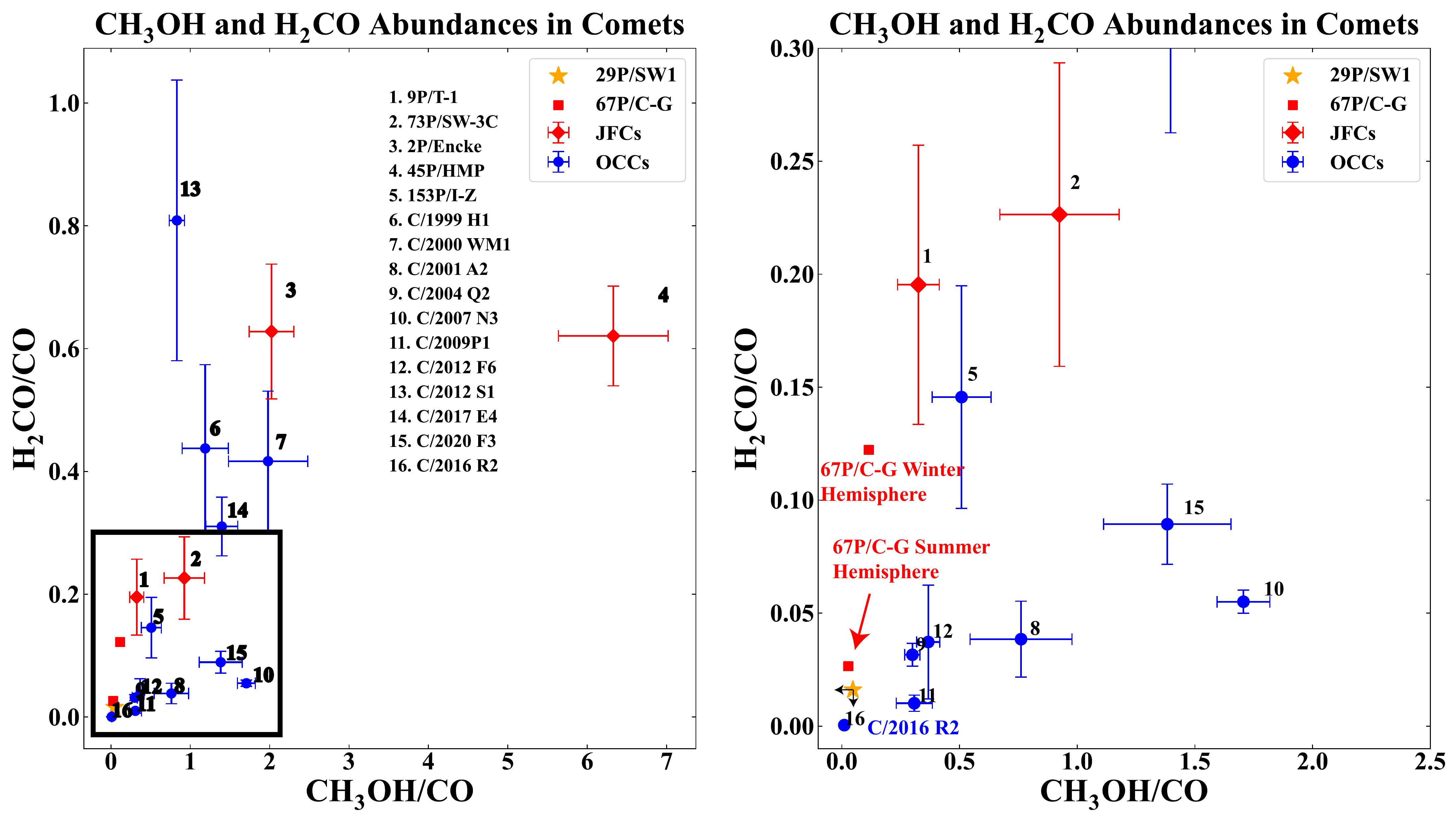}{\textwidth}{(B)}
	}
\caption{\textbf{(A).} Relative hypervolatiles abundances in comets \citep{DelloRusso2016a,LeRoy2015,Biver2018,McKay2019}. \textbf{B}. Relative CH$_3$OH and H$_2$CO abundances in comets \citep{DelloRusso2016a,LeRoy2015,Biver2018,McKay2019,Cordiner2022}. Arrows indicate 3$\sigma$ upper limits. Measurements for 67P/Churyumov-Gerasimenko in its summer and winter hemispheres \citep{LeRoy2015} are indicated. 
\label{fig:compare}}
\end{figure*}

Figure~\ref{fig:compare} demonstrates the considerable differences in SW1's coma composition compared to comets measured in the inner solar system. All measurements are from ground-based, high resolution near-infrared spectroscopy \citep{DelloRusso2016a,McKay2019} with the exception of radio wavelength measurements of C/2016 R2 \citep[PanSTARRS;][]{Biver2018,Cordiner2022} and \textit{in situ} Rosetta measurements of 67P/Churyumov-Gerasimenko \citep{LeRoy2015}. Differences in observing and data analysis techniques must be kept in mind when comparing these data sets. Of particular interest is how C$_2$H$_6$ and CH$_4$ abundances in SW1 compare to those observed in OCCs vs. JFCs. CO, CH$_4$, and C$_2$H$_6$ have the highest volatility among molecules routinely measured in comets and may be the most sensitive to thermal evolutionary processing. Thus, testing for compositional differences in the hypervolatiles between JFCs and OCCs can help to disentangle primordial from evolutionary effects in present-day measured coma composition \citep{DelloRusso2016a,DiSanti2017,Roth2018,Roth2020}. Comparing the primitive Kuiper disk material preserved in Centaurs against that measured in JFCs may provide additional insights into potential evolutionary processing suffered by JFCs relative to OCCs.

Figure~\ref{fig:compare} demonstrates that C$_2$H$_6$ and CH$_4$ abundances in JFCs span the same range of values measured in OCCs, although comet 45P/Honda-Mrkos-Pajdu\u{s}\'{a}kov\'{a} (point \#3) may be an outlier among the population. Notwithstanding the small number statistics for JFC hypervolatile abundances compared to OCCs, our sensitive 3$\sigma$ upper limits on C$_2$H$_6$ and CH$_4$ in SW1 are more consistent with some OCCs than with JFCs in general, with ultra CO-rich C/2016 R2 \citep[PanSTARRS;][]{Biver2018,McKay2019,Cordiner2022} being the closest match to SW1, followed by C/1995 O1 (Hale-Bopp). Our upper limits on both C$_2$H$_6$/CO and CH$_4$/CO are lower than that reported in any measured JFC. On the other hand, 3$\sigma$ upper limits on SW1's H$_2$CO and CH$_3$OH content are similar to values in C/2016 R2 (PanSTARRS) and in 67P/Churyumov-Gerasimenko measured by Rosetta \citep{LeRoy2015}. The similarities between SW1 and C/2016 R2 continue, with HCN/CO = (0.12 $\pm$ 0.03)\% in SW1 \citep{Bockelee2022} and (0.004 $\pm$ 0.001)\% for C/2016 R2 \citep{Biver2018} among the lowest values measured in comets.

\subsection{Compositional Similarities of SW1 and C/2016 R2 and Opportunities for Future Studies}\label{subsec:r2}
The remarkable similarity in volatile composition between SW1 and C/2016 R2, two comets with dramatically different dynamical histories, may be indicative of the form in which the ices are preserved in the comet nucleus. Given the very low CH$_4$/CO and H$_2$O/CO ratios in C/2016 R2, \cite{Cordiner2022} hypothesized that CH$_4$ is more associated with the polar (H$_2$O) ices in the nucleus. At the ultra-low temperatures in SW1, it is possible that the CH$_4$ may remain trapped in the frozen, polar ice phase, whereas the apolar (CO-rich) phase experiences more outgassing. 

As small bodies migrating from the scattered Kuiper disk onto Jupiter-family comet orbits, understanding the composition of Centaurs may provide a key avenue to disentangling potential  compositional differences between OCCs and JFCs and revealing whether they are natal or acquired. That SW1 is perhaps more compositionally similar to C/2016 R2, the most anomalous OCC discovered to date, than to JFCs or the general OCC population is intriguing and puzzling. C/2016 R2 displayed a consistently anomalous composition in all detected volatiles (by at least a factor of 3) compared to average OCCs, and its high CO and N$_2$ abundances compared to more complex species may suggest that it formed in a region of the protoplanetary disk that was chemically inactive and shielded from photodissociation \citep{McKay2019}. Alternately, its high hypervolatile content may be explained if it was the fragment of a differentiated Kuiper belt body \citep{Biver2018}. Although C/2016 R2 was measured at relatively large \rh{} ($\sim$3 au) compared to most measured comets, it was still considerably closer to the Sun than SW1, and the effects of \rh{} must be considered. Finally, it is worth remembering that SW1 itself is highly unusual even among the enigmatic Centaur class (or comets in general) owing to its remarkable cycle of outbursts and activity, whose mechanisms are still not fully understood.   

Definitively detecting the full hypervolatile suite (CO, CH$_4$, C$_2$H$_6$, CO$_2$) in SW1 and comparing the material preserved in Centaurs vs. JFCs will likely require the unprecedented sensitivity of the most advanced facilities, such as JWST and ALMA. JWST and the upcoming ALMA Wideband Sensitivity Upgrade will also enable the characterization of molecular chemistry in ever more distant OCCs, removing the bias of comet studies towards smaller \rh{} and testing how Centaur composition compares to OCC coma chemistry at large \rh{}. Centaurs are also under consideration for spaceflight missions as recommended by the recent Planetary Science \& Astrobiology Decadal Survey \citep{2023Decadal}. A Centaur orbiter and lander would provide paradigm-challenging insights for Centaurs analogous to those delivered for comets by the Rosetta rendezvous mission to comet 67P/Churyumov-Gerasimenko.  We strongly advocate for such missions and for observing programs that leverage state-of-the-art facilities for small body studies.

\section{Conclusion}\label{sec:conclusion}
The 2021 October outburst of SW1 presented an extraordinary opportunity to characterize the relative abundances of its volatiles across multiple wavelengths. Our nearly simultaneous IRTF and APEX/nFLASH230 observations provided highly consistent measures of CO production during the outburst, enabled sensitive upper limits on the abundances of trace species, gave measurements of gas expansion velocity crucial to accurate calculations of molecular production rates, and facilitated long-term monitoring to place the outburst into context with quiescent activity in SW1. Our results highlight the dramatic differences in coma composition between SW1 and comets measured in the inner solar system, and suggest that Centaurs may preserve material more similar to that found in some OCCs than in JFCs. Our nearly simultaneous near-infrared and radio measurements demonstrate the superb synergy that multi-wavelength measurements harness, with each component providing highly complementary science that cannot be achieved with either wavelength alone.

Unlocking the nature of primitive Kuiper disk material preserved within Centaurs will require characterizing the coma chemistry in a statistically significant number of these enigmatic objects. Our results, combined with the inherently faint nature of Centaurs, highlight the role that the most sensitive current and planned facilities, such as ALMA, JWST, and upcoming Extremely Large Telescopes (ELT's) will play in these efforts.

\acknowledgments
Data for this study were obtained at the NASA Infrared Telescope Facility (IRTF), operated by the University of Hawaii under contract NNH14CK55B with the National Aeronautics and Space Administration. We are most fortunate to the have the opportunity to conduct observations from Maunakea, and recognize the very significant cultural role and reverence that the summit of Maunakea has always had within the indigenous community. This publication is based on data acquired with the Atacama Pathfinder Experiment (APEX) under programme ID O-0108.F-0323A and O-0109.F-9308A. APEX is a collaboration between the Max-Planck-Institut fur Radioastronomie, the European Southern Observatory, and the Onsala Space Observatory. Swedish observations on APEX are supported through Swedish Research Council grant No 2017-00648. The APEX data were reduced thanks to the use of the \texttt{GIDLAS/CLASS} software (\url{http://www.iram.fr/IRAMFR/GILDAS}). This work was supported by the Planetary Science Division Internal Scientist Funding Program through the Fundamental Laboratory Research (FLaRe) work package (NXR, SNM, MAC, SBC, SF, GLV, MAD, AJM). Part of this research was carried out at the Jet Propulsion Laboratory, California Institute of Technology, under a contract with the National Aeronautics and Space Administration (80NM0018D0004). BPB acknowledges support from the National Science Foundation (NSF AST-2009398).

\appendix
\section{Correction for Optical Depth in the PSG}\label{sec:odpsg}
Specific fluorescence emissions originate from a myriad of pump and cascade processes. Considering the high-energy of the solar pumping flux, comprehensive high-energy pump linelists including billions of transitions are required. Treatment of opacity and full radiative transfer employing such large databases, in particular for non-resonant fluorescence, can therefore be extremely challenging.

In the PSG correction for optical depth we kept track of the associated line intensities that led to the specific emission for every $g$-factor. We then computed a weighted ``representative'' line intensity $S_p$ [cm$^{-1}$/(molecule\,cm$^{-2}$)] and documented for each $g$-factor its weighted pump intensity. The weight is defined based on the pump intensity ($g_{lu}$) divided by the pump line frequency (cm$^{-1}$). The inclusion of the frequency in the weight originates from the fact that the calculation of the opacity employs the emission frequency, not the pump frequency. 

In the case of a non-resonant emission for $v=1\rightarrow0$ originating from a pump of $v=0\rightarrow2$ (later cascading to $v=1$), the line width at the pump is twice as big in wavenumbers (cm$^{-1}$) and the opacity lower by a factor of two for the same line intensity. For instance, for the ro-vibrational P2 transition (\Ju{}=1 to \Ju{}=2) of CO, the pumps to \Ju{}=1 originate from the lines R0 (\Ju{}=0 to \Ju{}=1) and P2 (\Ju{}=2 to \Ju{}=1), while the effective line intensity is calculated as
\begin{equation}
S_p = f_{ul,P2}\frac{(g_{lu,R0}S_{lu,R0}/f_{lu,R0}+g_{lu,P2}S_{lu,P2}/f_{lu,P2})}{(g_{lu,R0}+g_{lu,P2})}
\end{equation}

where $g_{lu}$ is the pump $g$-factor (s$^{-1}$), $f_{lu}$ is the frequency of the line (cm$^{-1}$) at the pump, $f_{ul}$ is the emission frequency (cm$^{-1}$), and $S_{lu}$ is the line intensity [cm$^{-1}$/(molecule\,cm$^{-2}$)] at the specific rotational temperature, population state, and heliocentric velocity. This was generalized in the fluorescence model to allow for complex non-resonant cascades by computing $S_p$ following branching ratios and weights as done for the emission $g$-factor \citep[see][for further details]{Villanueva2022}.

For low Sun-Comet-Observer phase angles, the integrated column density as measured by the observer also describes the column density of the incident solar flux. The average linewidth of the pump ($w_p$, cm$^{-1}$) can be computed as 
$w_p = (2v_pf_{ul})/{c}$, where $v_p$ is the expansion velocity (m\,s$^{-1}$), $f_{ul}$ is the emission frequency (cm$^{-1}$), and $c$ is the speed of light. The integrated opacity across the pump can be calculated as $\tau_p=N_{col}S_p/v_p$, where $N_{col}$ is the integrated column density (molecules\,cm$^{-2}$) along the line of sight. The transmittance at the end of the column is $e^{-\tau_p}$ at the frequency of the pump. For low opacities, there are few molecules attenuating the solar flux at the pump frequency and the pumps are optically thin. As the opacity increases across the column for the solar pump, the observer only receives radiation for the molecules up to $\tau_p<1$, and therefore the expected fluorescence efficiency can be approximated as 

\begin{equation}
g_{thick} = \frac{1-e^{-\tau_p}}{\tau_p}g_{thin}
\end{equation}

This is a first order correction to a complex problem and is only valid for low solar phases. Nevertheless, by documenting the opacity at the pump, the PSG can provide guidance to the user on the level of opacity in the synthetic spectra. Applying the correction, the optically thin column density (before convolution with the seeing PSF) is related to the optically thin column density ($N_i$) and optically thin $g$-factors ($g_i$) as

\begin{equation}
N_{thick} = \frac{\sum_i \frac{(1-e^{-\tau_i})}{\tau_i}g_i}{\sum_i g_i}N_{thin}
\end{equation}

\bibliography{29P}{}

\begin{thebibliography}{}
\expandafter\ifx\csname natexlab\endcsname\relax\def\natexlab#1{#1}\fi
\providecommand{\url}[1]{\href{#1}{#1}}
\providecommand{\dodoi}[1]{doi:~\href{http://doi.org/#1}{\nolinkurl{#1}}}
\providecommand{\doeprint}[1]{\href{http://ascl.net/#1}{\nolinkurl{http://ascl.net/#1}}}
\providecommand{\doarXiv}[1]{\href{https://arxiv.org/abs/#1}{\nolinkurl{https://arxiv.org/abs/#1}}}

\bibitem[{{A'Hearn} {et~al.}(1995){A'Hearn}, {Millis}, {Schleicher}, {Osip}, \&
  {Birch}}]{AHearn1995}
{A'Hearn}, M.~F., {Millis}, R.~L., {Schleicher}, D.~G., {Osip}, D.~J., \&
  {Birch}, V.~P. 1995, Icarus, 118, 223

\bibitem[{{Biver} {et~al.}(2018){Biver}, {Bockel{\'e}e-Morvan}, {Paubert},
  {Moreno}, {Crovisier}, {Boissier}, {Bertrand}, {Boussier}, {Kugel}, {McKay},
  {Dello Russo}, \& {DiSanti}}]{Biver2018}
{Biver}, N., {Bockel{\'e}e-Morvan}, D., {Paubert}, G., {et~al.} 2018, A\&A,
  619, A127, \dodoi{10.1051/0004-6361/201833449}

\bibitem[{{Bockelee-Morvan}(1987)}]{Bockelee1987}
{Bockelee-Morvan}, D. 1987, A\&, 181, 169

\bibitem[{{Bockel{\'e}e-Morvan} \& {Biver}(2017)}]{Bockelee2017}
{Bockel{\'e}e-Morvan}, D., \& {Biver}, N. 2017, PTRSA, 375, 20160252

\bibitem[{{Bockel{\'e}e-Morvan} {et~al.}(2010){Bockel{\'e}e-Morvan},
  {Boissier}, {Biver}, \& {Crovisier}}]{Bockelee2010}
{Bockel{\'e}e-Morvan}, D., {Boissier}, J., {Biver}, N., \& {Crovisier}, J.
  2010, Icarus, 210, 898, \dodoi{10.1016/j.icarus.2010.07.005}

\bibitem[{{Bockel{\'e}e-Morvan} {et~al.}(2004){Bockel{\'e}e-Morvan},
  {Crovisier}, {Mumma}, \& {Weaver}}]{Bockelee2004}
{Bockel{\'e}e-Morvan}, D., {Crovisier}, J., {Mumma}, M.~J., \& {Weaver}, H.~A.
  2004, in Comets II, ed. H.~U. {Keller} \& H.~A. {Weaver} (University of
  Arizona Press), 391

\bibitem[{{Bockel{\'e}e-Morvan} {et~al.}(2022){Bockel{\'e}e-Morvan}, {Biver},
  {Schambeau}, {Crovisier}, {Opitom}, {de Val Borro}, {Lellouch}, {Hartogh},
  {Vandenbussche}, {Jehin}, {Kidger}, {K{\"u}ppers}, {Lis}, {Moreno},
  {Szutowicz}, \& {Zakharov}}]{Bockelee2022}
{Bockel{\'e}e-Morvan}, D., {Biver}, N., {Schambeau}, C.~A., {et~al.} 2022,
  A\&A, 664, A95

\bibitem[{{Boissier} {et~al.}(2007){Boissier}, {Bockel{\'e}e-Morvan}, {Biver},
  {Crovisier}, {Despois}, {Marsden}, \& {Moreno}}]{Boissier2007}
{Boissier}, J., {Bockel{\'e}e-Morvan}, D., {Biver}, N., {et~al.} 2007, A\&A,
  475, 1131

\bibitem[{Bonev(2005)}]{Bonev2005}
Bonev, B.~P. 2005, phdthesis, The University of Toledo

\bibitem[{{Bonev} {et~al.}(2008){Bonev}, {Mumma}, {Radeva}, {DiSanti}, {Gibb},
  \& {Villanueva}}]{Bonev2008}
{Bonev}, B.~P., {Mumma}, M.~J., {Radeva}, Y.~L., {et~al.} 2008, ApJL, 680, L61

\bibitem[{Bonev {et~al.}(2017)Bonev, Villanueva, DiSanti, Boehnhardt, Lippi,
  Gibb, Paganini, \& Mumma}]{Bonev2017}
Bonev, B.~P., Villanueva, G.~L., DiSanti, M.~A., {et~al.} 2017, The
  Astronomical Journal, 153, 241

\bibitem[{{Cochran} \& {Cochran}(1991)}]{Cochran1991}
{Cochran}, A.~L., \& {Cochran}, W.~D. 1991, Icarus, 90, 172,
  \dodoi{10.1016/0019-1035(91)90077-7}

\bibitem[{{Cordiner} {et~al.}(2022){Cordiner}, {Coulson}, {Garcia-Berrios},
  {Qi}, {Lique}, {Zo{\l}towski}, {de Val-Borro}, {Kuan}, {Ip}, {Mairs}, {Roth},
  {Charnley}, {Milam}, {Tseng}, \& {Chuang}}]{Cordiner2022}
{Cordiner}, M.~A., {Coulson}, I.~M., {Garcia-Berrios}, E., {et~al.} 2022, ApJ,
  929, 38

\bibitem[{Dello~Russo {et~al.}(2016a)Dello~Russo, Kawakita, Jr., \&
  Weaver~H.}]{DelloRusso2016a}
Dello~Russo, N., Kawakita, H., Jr., V. R.~J., \& Weaver~H., A. 2016a, Icarus,
  278, 301

\bibitem[{{Denis-Alpizar} {et~al.}(2018){Denis-Alpizar}, {Stoecklin},
  {Guilloteau}, \& {Dutrey}}]{Denis-alpizar2018}
{Denis-Alpizar}, O., {Stoecklin}, T., {Guilloteau}, S., \& {Dutrey}, A. 2018,
  MNRAS, 478, 1811, \dodoi{10.1093/mnras/sty1177}

\bibitem[{DiSanti {et~al.}(2017)DiSanti, Bonev, Dello~Russo, Vervack, Gibb,
  Roth, McKay, \& Kawakita}]{DiSanti2017}
DiSanti, M.~A., Bonev, B.~P., Dello~Russo, N., {et~al.} 2017, AJ, 154, 246

\bibitem[{DiSanti {et~al.}(2006)DiSanti, Bonev, Magee-Sauer, Dello~Russo,
  Mumma, Reuter, \& Villanueva}]{DiSanti2006}
DiSanti, M.~A., Bonev, B.~P., Magee-Sauer, K., {et~al.} 2006, ApJ, 650, 470

\bibitem[{DiSanti {et~al.}(2001)DiSanti, Mumma, Dello~Russo, \&
  Magee-Sauer}]{DiSanti2001}
DiSanti, M.~A., Mumma, M.~J., Dello~Russo, N., \& Magee-Sauer, K. 2001, Icarus,
  153, 361

\bibitem[{DiSanti {et~al.}(2003)DiSanti, Mumma, Dello~Russo, Magee-Sauer, \&
  Griep}]{DiSanti2003}
DiSanti, M.~A., Mumma, M.~J., Dello~Russo, N., Magee-Sauer, K., \& Griep, D.~M.
  2003, JGRP, 108, 5061

\bibitem[{DiSanti {et~al.}(2014)DiSanti, Villanueva, Paganini, Bonev, Keane,
  Meech, \& Mumma}]{DiSanti2014}
DiSanti, M.~A., Villanueva, G.~L., Paganini, L., {et~al.} 2014, Icarus, 228,
  167

\bibitem[{DiSanti {et~al.}(2016)DiSanti, Bonev, Gibb, Paganini, Villanueva,
  Mumma, Keane, Blake, Dello~Russo, Meech, Vervack, \& McKay}]{DiSanti2016}
DiSanti, M.~A., Bonev, B.~P., Gibb, E.~L., {et~al.} 2016, ApJ, 820, 20

\bibitem[{Faggi {et~al.}(2018)Faggi, Villanueva, Mumma, \&
  Paganini}]{Faggi2018}
Faggi, S., Villanueva, G.~L., Mumma, M.~J., \& Paganini, L. 2018, The
  Astronomical Journal, 156, 68

\bibitem[{{Gunnarsson} {et~al.}(2008){Gunnarsson}, {Bockel{\'e}e-Morvan},
  {Biver}, {Crovisier}, \& {Rickman}}]{Gunnarsson2008}
{Gunnarsson}, M., {Bockel{\'e}e-Morvan}, D., {Biver}, N., {Crovisier}, J., \&
  {Rickman}, H. 2008, A\&A, 484, 537, \dodoi{10.1051/0004-6361:20078069}

\bibitem[{{G{\"u}sten} {et~al.}(2006){G{\"u}sten}, {Nyman}, {Schilke},
  {Menten}, {Cesarsky}, \& {Booth}}]{Gusten2006}
{G{\"u}sten}, R., {Nyman}, L.~{\r{A}}., {Schilke}, P., {et~al.} 2006, A\&A,
  454, L13

\bibitem[{{Harrington Pinto} {et~al.}(2022){Harrington Pinto}, {Womack},
  {Fernandez}, \& {Bauer}}]{Harrington2022}
{Harrington Pinto}, O., {Womack}, M., {Fernandez}, Y., \& {Bauer}, J. 2022,
  PSJ, 3, 247, \dodoi{10.3847/PSJ/ac960d}

\bibitem[{{Haser}(1957)}]{Haser1957}
{Haser}, L. 1957, BSRSL, 43, 740

\bibitem[{{Huebner} \& {Mukherjee}(2015)}]{Huebner2015}
{Huebner}, W.~F., \& {Mukherjee}, J. 2015, P\&SS, 106, 11,
  \dodoi{10.1016/j.pss.2014.11.022}

\bibitem[{{Hughes}(1990)}]{Hughes1990}
{Hughes}, D.~W. 1990, QJRAS, 31, 69

\bibitem[{{Ivanova} {et~al.}(2016){Ivanova}, {Luk`yanyk}, {Kiselev},
  {Afanasiev}, {Picazzio}, {Cavichia}, {de Almeida}, \&
  {Andrievsky}}]{Ivanova2016}
{Ivanova}, O.~V., {Luk`yanyk}, I.~V., {Kiselev}, N.~N., {et~al.} 2016, P\&SS,
  121, 10, \dodoi{10.1016/j.pss.2015.12.001}

\bibitem[{{Jewitt}(2009)}]{Jewitt2009}
{Jewitt}, D. 2009, AJ, 137, 4296, \dodoi{10.1088/0004-6256/137/5/4296}

\bibitem[{{Korsun} {et~al.}(2008){Korsun}, {Ivanova}, \&
  {Afanasiev}}]{Korsun2008}
{Korsun}, P.~P., {Ivanova}, O.~V., \& {Afanasiev}, V.~L. 2008, Icarus, 198,
  465, \dodoi{10.1016/j.icarus.2008.08.010}

\bibitem[{Le~Roy {et~al.}(2015)Le~Roy, Altwegg, Balsiger, Berthelier, Bieler,
  Briois, Calmonte, Combi, De~Keyser, Dhooghe, Fiethe, Fuselier, Gasc, Gombosi,
  H{\"a}ssig, J{\"a}ckel, Rubin, \& Tzou}]{LeRoy2015}
Le~Roy, L., Altwegg, K., Balsiger, H., {et~al.} 2015, Astronomy \&
  Astrophysics, 583, A1

\bibitem[{{McKay} {et~al.}(2019){McKay}, {DiSanti}, Kelley, {Knight}, {Womack},
  {Wierzchos}, {Harrington Pinto}, {Bonev}, {Villanueva}, {Dello Russo},
  {Cochran}, {Biver}, {Bauer}, {Vervack}, {Gibb}, {Roth}, \&
  {Kawakita}}]{McKay2019}
{McKay}, A.~J., {DiSanti}, M.~A., Kelley, M. S.~P., {et~al.} 2019, AJ, 158, 128

\bibitem[{{Meech} {et~al.}(2009){Meech}, {Pittichov{\'a}}, {Bar-Nun},
  {Notesco}, {Laufer}, {Hainaut}, {Lowry}, {Yeomans}, \& {Pitts}}]{Meech2009}
{Meech}, K.~J., {Pittichov{\'a}}, J., {Bar-Nun}, A., {et~al.} 2009, Icarus,
  201, 719

\bibitem[{{Miles} \& {Mission29P}(2021)}]{Miles2021}
{Miles}, R., \& {Mission29P}. 2021, {Centaur Comet 29P/Schwassmann-Wachmann
  2021-2022 Apparition}.
\newblock
  \url{https://britastro.org/section_information_/comet-section-overview/mission-29p-centaur-comet-observing-campaign}

\bibitem[{Mumma \& Charnley(2011)}]{Mumma2011a}
Mumma, M.~J., \& Charnley, S.~B. 2011, ARA\&A, 49, 471

\bibitem[{{National Academies of Sciences} {et~al.}(2022){National Academies of
  Sciences}, {Engineering}, \& {Medicine}}]{2023Decadal}
{National Academies of Sciences}, {Engineering}, \& {Medicine}. 2022, {Origins,
  Worlds, and Life: A Decadal Strategy for Planetary Science and Astrobiology
  2023-2032} (Washington, DC: The National Academies Press),
  \dodoi{10.17226/26522}

\bibitem[{Ootsubo {et~al.}(2012)Ootsubo, Kawakita, Hamada, Kobayashi,
  Yamaguchi, Usui, Nakagawa, Ueno, Ishiguro, Sekiguchi, Watanabe, Sakon,
  Shimonishi, \& Onaka}]{Ootsubo2012}
Ootsubo, T., Kawakita, H., Hamada, S., {et~al.} 2012, The Astrophysical
  Journal, 752, 15

\bibitem[{{Paganini} {et~al.}(2013){Paganini}, {Mumma}, {Boehnhardt},
  {DiSanti}, {Villanueva}, {Bonev}, {Lippi}, {K{\"a}ufl}, \&
  {Blake}}]{Paganini2013}
{Paganini}, L., {Mumma}, M.~J., {Boehnhardt}, H., {et~al.} 2013, ApJ, 766, 100,
  \dodoi{10.1088/0004-637X/766/2/100}

\bibitem[{{Prialnik} \& {Bar-Nun}(1987)}]{Prialnik1987}
{Prialnik}, D., \& {Bar-Nun}, A. 1987, ApJ, 313, 893, \dodoi{10.1086/165029}

\bibitem[{{Rabli} \& {Flower}(2010)}]{Rabli2010}
{Rabli}, D., \& {Flower}, D.~R. 2010, MNRAS, 406, 95,
  \dodoi{10.1111/j.1365-2966.2010.16671.x}

\bibitem[{Radeva {et~al.}(2010)Radeva, Mumma, Bonev, DiSanti, Villanueva,
  Magee-Sauer, Gibb, \& Weaver}]{Radeva2010}
Radeva, Y.~L., Mumma, M.~J., Bonev, B.~P., {et~al.} 2010, Icarus, 206, 764

\bibitem[{Rayner {et~al.}(2012)Rayner, Bond, Bonnet, Jaffe, Muller, \&
  Tokunaga}]{Rayner2012}
Rayner, J., Bond, T., Bonnet, M., {et~al.} 2012, Proc. SPIE, 8446, 84462C

\bibitem[{Rayner {et~al.}(2016)Rayner, Tokunaga, Jaffe, Bonnet, Ching,
  Connelley, Kokubun, Lockhart, \& Warmbier}]{Rayner2016}
Rayner, J., Tokunaga, A., Jaffe, D., {et~al.} 2016, Proc. SPIE, 9908, 990884

\bibitem[{Roth {et~al.}(2018)Roth, Gibb, Bonev, DiSanti, Dello~Russo, Vervack,
  McKay, \& Kawakita}]{Roth2018}
Roth, N.~X., Gibb, E.~L., Bonev, B.~P., {et~al.} 2018, AJ, 156, 251

\bibitem[{{Roth} {et~al.}(2020){Roth}, {Gibb}, {Bonev}, {DiSanti}, {Dello
  Russo}, {McKay}, {Vervack}, {Kawakita}, {Saki}, {Biver},
  {Bockel{\'e}e-Morvan}, {Feaga}, {Fougere}, {Cochran}, {Combi}, \&
  {Shou}}]{Roth2020}
{Roth}, N.~X., {Gibb}, E.~L., {Bonev}, B.~P., {et~al.} 2020, AJ, 159, 42

\bibitem[{{Roth} {et~al.}(2021){Roth}, {Bonev}, {DiSanti}, {Dello Russo},
  {McKay}, {Gibb}, {Saki}, {Khan}, {Vervack}, {Kawakita}, {Cochran}, {Biver},
  {Cordiner}, {Crovisier}, {Jehin}, \& {Weaver}}]{Roth2021b}
{Roth}, N.~X., {Bonev}, B.~P., {DiSanti}, M.~A., {et~al.} 2021, PSJ, 2, 54,
  \dodoi{10.3847/PSJ/abd706/54}

\bibitem[{{Sch{\"o}ier} {et~al.}(2005){Sch{\"o}ier}, {van der Tak}, {van
  Dishoeck}, \& {Black}}]{Schoier2005}
{Sch{\"o}ier}, F.~L., {van der Tak}, F.~F.~S., {van Dishoeck}, E.~F., \&
  {Black}, J.~H. 2005, \aap, 432, 369, \dodoi{10.1051/0004-6361:20041729}

\bibitem[{{Stern}(2003)}]{Stern2003}
{Stern}, S.~A. 2003, Nature, 424, 639

\bibitem[{{Villanueva} {et~al.}(2022){Villanueva}, {Liuzzi}, {Faggi},
  {Protopapa}, {Kofman}, {Fauchez}, {Stone}, \& {Mandell}}]{Villanueva2022}
{Villanueva}, G., {Liuzzi}, G., {Faggi}, S., {et~al.} 2022, in Fundamentals of
  the Planetary Spectrum Generator 2022 Edition, Vol.~1, 43--63

\bibitem[{Villanueva {et~al.}(2009)Villanueva, Mumma, Bonev, DiSanti, Gibb,
  Boehnhardt, \& Lippi}]{Villanueva2009}
Villanueva, G.~L., Mumma, M.~J., Bonev, B.~P., {et~al.} 2009, ApJL, 690, L5

\bibitem[{{Villanueva} {et~al.}(2012b){Villanueva}, {Mumma}, {Bonev}, {Novak},
  {Barber}, \& {DiSanti}}]{Villanueva2012b}
{Villanueva}, G.~L., {Mumma}, M.~J., {Bonev}, B.~P., {et~al.} 2012b, JQSRT,
  113, 202

\bibitem[{Villanueva {et~al.}(2011a)Villanueva, Mumma, DiSanti, Bonev, Gibb,
  Magee-Sauer, Blake, \& Salyk}]{Villanueva2011a}
Villanueva, G.~L., Mumma, M.~J., DiSanti, M.~A., {et~al.} 2011a, Icarus, 216,
  227

\bibitem[{Villanueva {et~al.}(2008)Villanueva, Mumma, Novak, \&
  Hewagama}]{Villanueva2008}
Villanueva, G.~L., Mumma, M.~J., Novak, R.~E., \& Hewagama, T. 2008, Icarus,
  195, 34

\bibitem[{{Villanueva} {et~al.}(2018){Villanueva}, {Smith}, {Protopapa},
  {Faggi}, \& {Mandell}}]{Villanueva2018}
{Villanueva}, G.~L., {Smith}, M.~D., {Protopapa}, S., {Faggi}, S., \&
  {Mandell}, A.~M. 2018, JQSRT, 217, 86

\bibitem[{{Villanueva} {et~al.}(2013b){Villanueva}, {Mumma}, {Novak}, {Radeva},
  {K{\"a}ufl}, {Smette}, {Tokunaga}, {Khayat}, {Encrenaz}, \&
  {Hartogh}}]{Villanueva2013b}
{Villanueva}, G.~L., {Mumma}, M.~J., {Novak}, R.~E., {et~al.} 2013b, Icarus,
  223, 11, \dodoi{10.1016/j.icarus.2012.11.013}

\bibitem[{{Wierzchos} \& {Womack}(2020)}]{Wierzchos2020}
{Wierzchos}, K., \& {Womack}, M. 2020, AJ, 159, 136,
  \dodoi{10.3847/1538-3881/ab6e68}

\bibitem[{{Wiesenfeld} \& {Faure}(2013)}]{Wiesenfeld2013}
{Wiesenfeld}, L., \& {Faure}, A. 2013, MNRAS, 432, 2573,
  \dodoi{10.1093/mnras/stt616}

\end{thebibliography}
\bibliographystyle{aasjournal}



\end{document}